\documentclass{egpubl}
\usepackage{eurovis2022}

\JournalSubmission    
\usepackage[T1]{fontenc}
\usepackage{dfadobe}  

\usepackage{cite}  
\BibtexOrBiblatex
\electronicVersion
\PrintedOrElectronic
\ifpdf \usepackage[pdftex]{graphicx} \pdfcompresslevel=9
\else \usepackage[dvips]{graphicx} \fi

\usepackage{egweblnk}

\usepackage{egweblnk}
\usepackage[dvipsnames]{xcolor}
\usepackage{lipsum}
\usepackage{todonotes}
\usepackage{pdfpages}

\usepackage{url}

\definecolor{myblue}{rgb}{0.0156, 0.2431, 0.5843}
\definecolor{gray}{rgb}{0.7,0.7,0.7}
\definecolor{orange}{rgb}{1.0,0.5,0.0}
\newcommand*{\rv}{\textcolor{black}}
\newcommand*{\chan}{\textcolor{black}}

\makeatletter
\newcommand{\printfnsymbol}[1]{%
}
\makeatother

\title[Nested Papercrafts for Anatomical and Biological Edutainment]%
      {Nested Papercrafts for Anatomical and Biological Edutainment}

\author[M.~Schindler \& T.~Korpitsch \& R.~G.~Raidou \& H.-Y.~Wu]
{\parbox{\textwidth}{\centering 
M.~Schindler\thanks{Equal contribution}
$^{1}$,
T.~Korpitsch\footnotemark[1]
$^{1}$\orcid{0000-0003-1919-467X},
R.~G.~Raidou
$^{1}$\orcid{0000-0003-2468-0664},
and H.-Y.~Wu
$^{1,2}$\orcid{0000-0003-1028-0010} 
}
\\
{\parbox{\textwidth}{\centering 
$^1$TU Wien, Austria 
$^2$St. P{\"o}lten University of Applied Sciences, Austria
       }
}
}

\usepackage{enumerate}                
\usepackage[shortlabels]{enumitem}

\usepackage{lineno}
\EGpagenumber

\begin{document}


\teaser{
\vspace{-35pt}
\centering{
 \setlength{\tabcolsep}{0pt}
 \begin{tabular}{ccccc}
    \includegraphics[width=0.20\linewidth]{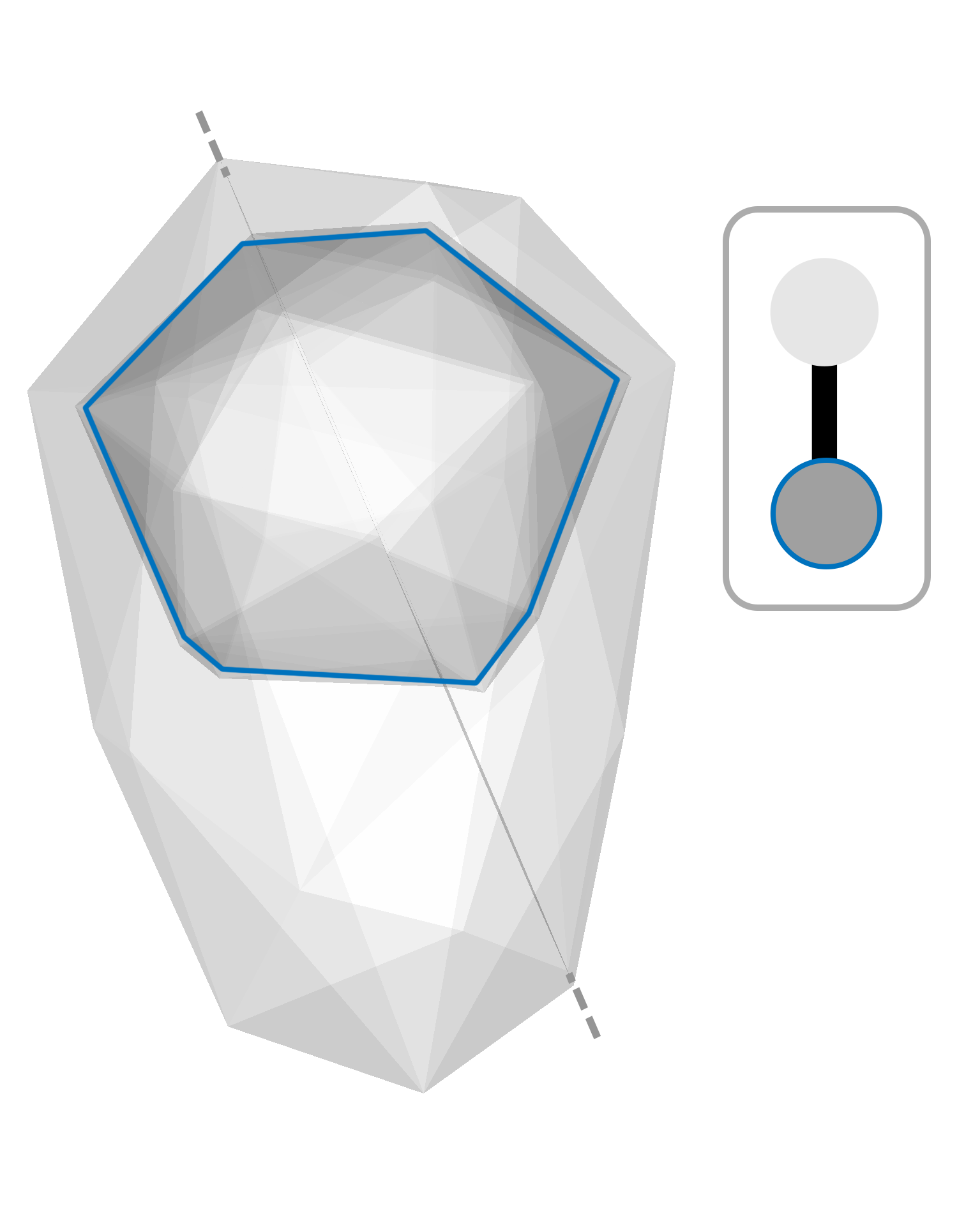} &
    \includegraphics[width=0.18\linewidth]{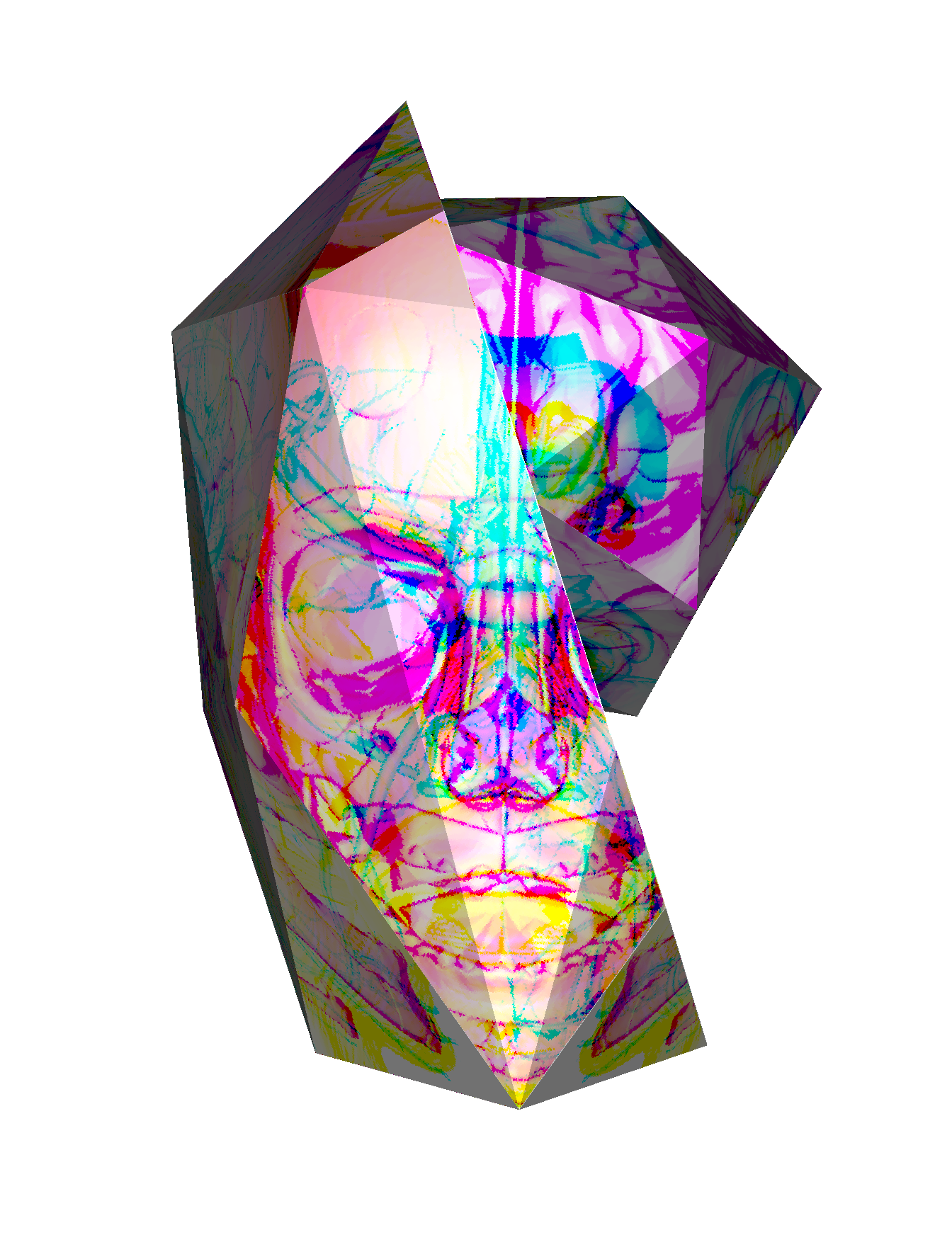} &
    \includegraphics[width=0.18\linewidth]{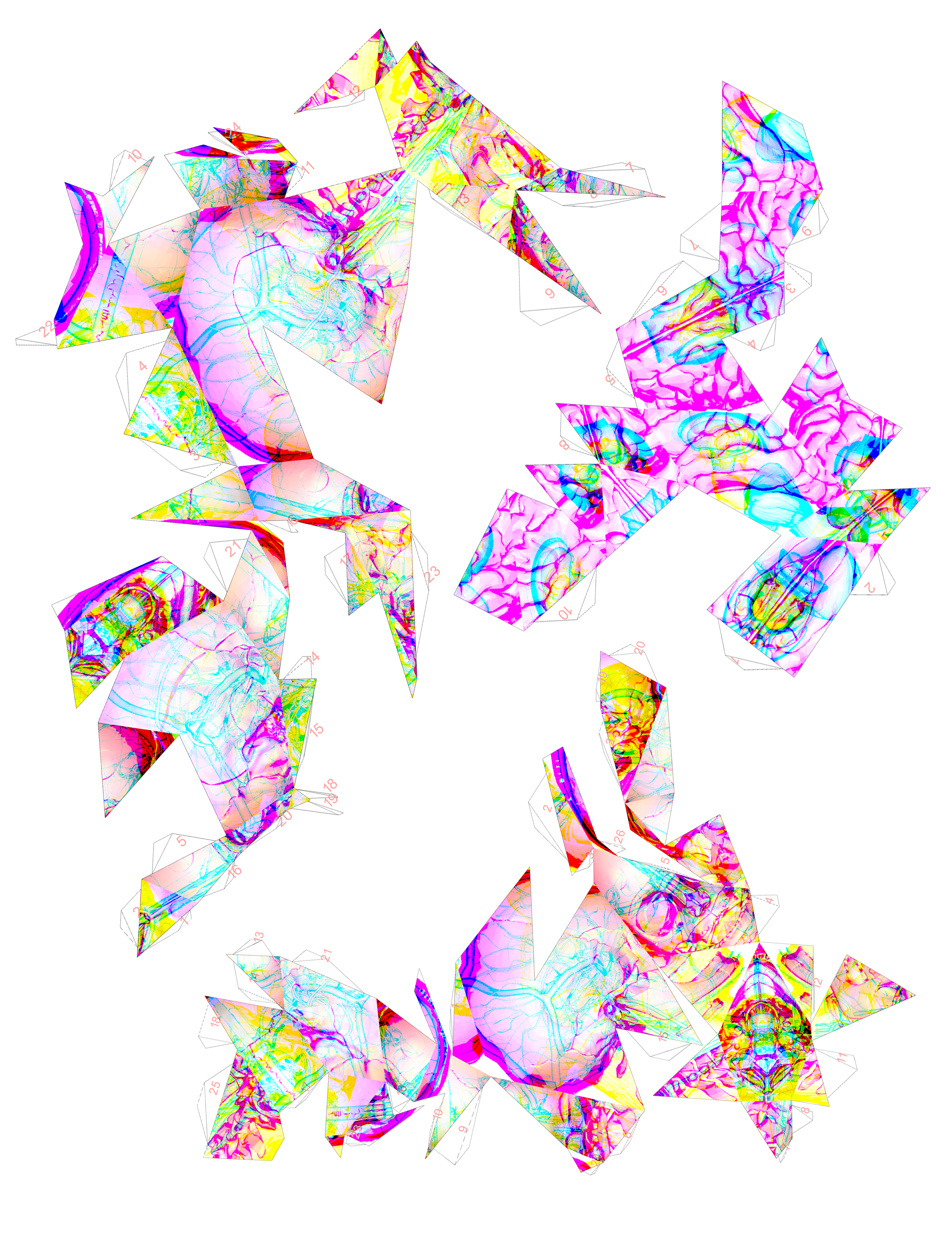} &
    \includegraphics[width=0.22\linewidth]{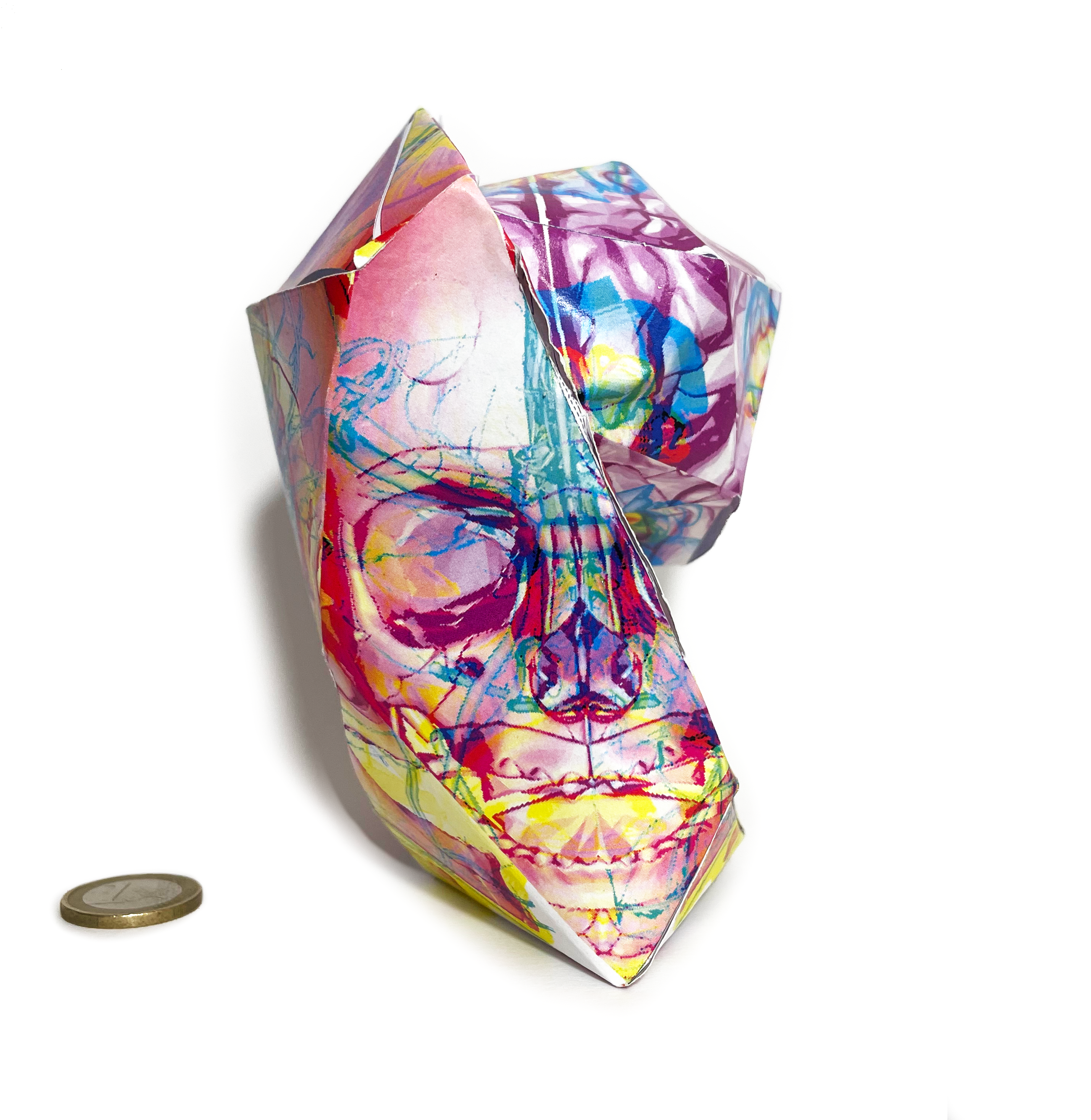} &
    \includegraphics[width=0.18\linewidth]{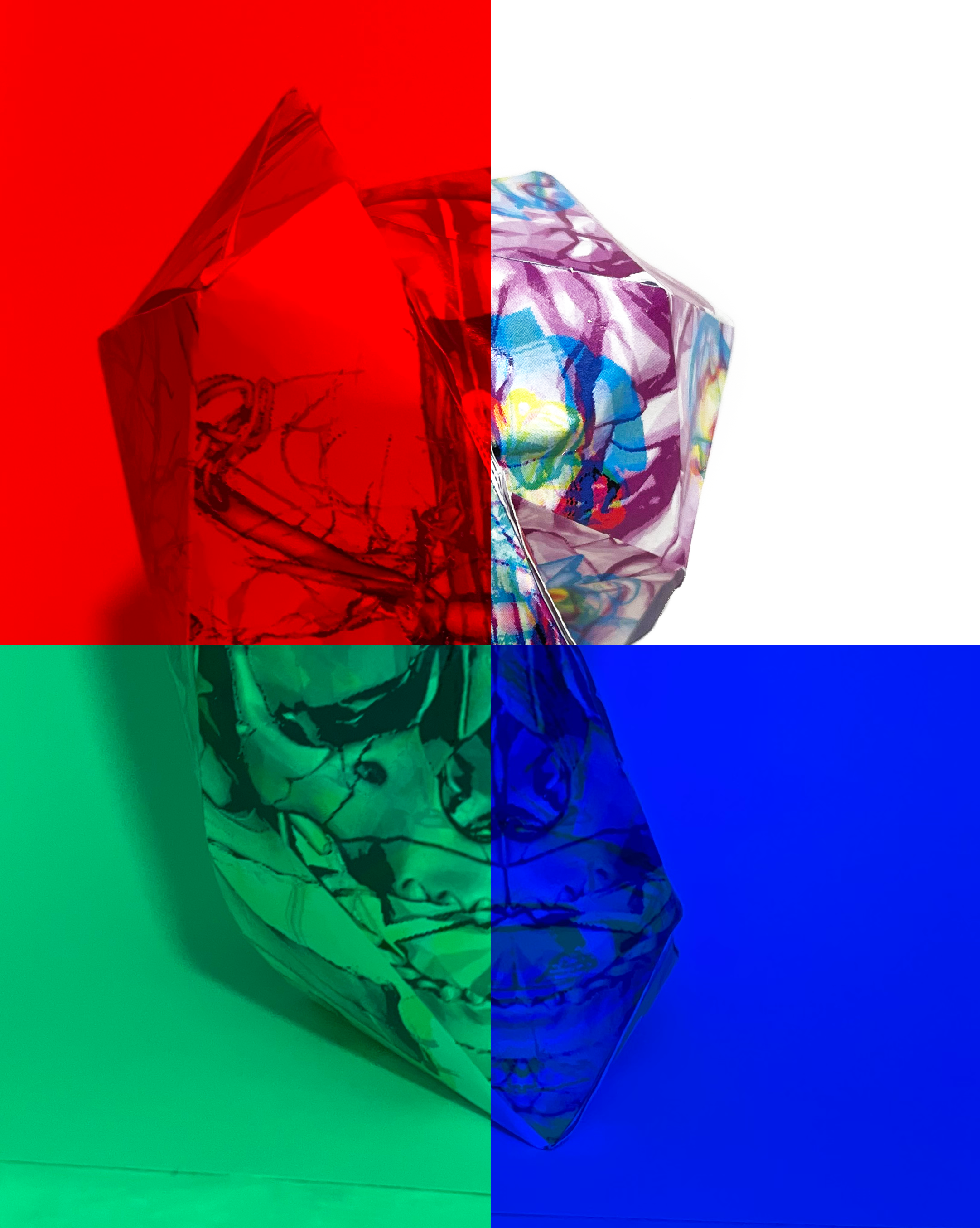} \\
   (a) & 
   (b) &
   (c) &
   (d) &
   (e) \\
   \\
 \end{tabular}
}
\caption{An example of \rv{a} nested papercraft for the exploration of the structures of a human head dataset, with two levels. The outer level represents the outer, visible part of the head and the inner level represents the brain. We show (a) the calculation of the optimal cut for the nested configuration, (b) the texture rendering for the papermeshes of the two levels, (c) the unfolding of the textures onto single 2D patches, (d) the assembly of the nested papercraft, and (e) the use of colored filters to reveal distinct features on the textured papermeshes.}
\label{fig:teaser}
}

\maketitle
  \begin{abstract} 
\rv{In this paper, we present a new workflow for the computer-aided generation of physicalizations, addressing nested configurations in anatomical and biological structures.}
\rv{Physicalizations are an important component of anatomical and biological education and edutainment.
However, existing approaches have mainly revolved around creating data sculptures through digital fabrication. 
Only a few recent works proposed computer-aided pipelines for generating sculptures, \rv{such as papercrafts,} with affordable and readily available materials.
Papercraft generation remains a challenging topic by itself. 
Yet, anatomical and biological applications pose additional challenges, such as reconstruction complexity and insufficiency to account for multiple, nested structures---often present in anatomical and biological structures. }
\rv{Our workflow comprises the following steps:
\emph{(i)} define the nested configuration of the model and detect its levels, 
\emph{(ii)} calculate the viewpoint that provides optimal, unobstructed views on inner levels,
\emph{(iii)} perform cuts on the outer levels to reveal the inner ones based on the viewpoint selection,
\emph{(iv)} estimate the stability of the cut papercraft to ensure a reliable outcome,
\emph{(v)} generate textures at each level, as a smart visibility mechanism that provides additional information on the inner structures, and
\emph{(vi)} unfold each textured mesh guaranteeing reconstruction.
Our novel approach exploits the interactivity of nested papercraft models for edutainment purposes. 
}


\end{abstract}  

\section{Introduction}
\label{sec:intro}


Visualization 
is recognized as an essential component of learning and education~\cite{gilbert2005visualization}. 
In the last few years, the impact of visualization in \textit{edutainment} (i.e., educational entertainment) has also been investigated---for example within the context of augmented~\cite{stefan2014ar} and virtual reality~\cite{cai2006bio}, or in comparison to serious games~\cite{charsky2010edutainment}. 
\textit{Physicalization} is a subdomain of visualization, which revolves around creating physical objects to represent and explore data~\cite{djavaherpour2021data}.
The field has recently seen a resurgence, also for education and edutainment purposes~\cite{perin2021students}. 
Generated physical objects are tangible and employ more senses than plain vision, to enhance our cognition and perception about displayed information~\cite{Embodiment, jansen2015opportunities}. 
Physicalizations provide a high degree of engagement, supporting understanding and memorability~\cite{jansen2013evaluating,SS15}, while also making the entire process interactive and entertaining. 

In education, physicalizations may be used either for \emph{demonstration and presentation} purposes with pre-made data models or for \emph{engaging students} in the learning processes by letting them handcraft their representations. 
In edutainment, 
physicalizations would rather focus on the latter. 
So far, the generation of physicalizations has mainly focused on \emph{digital fabrication methods} (e.g., 3D printing, laser cutting, moulding, etc.)~\cite{bickel2018state}, \emph{augmentations} (e.g., projections on passive props and augmented perspectives), and \emph{active approaches} (e.g., data sculptures that use materials to communicate data, robots, and screens)~\cite{djavaherpour2021data}.

In biology and anatomy education, physical models have always been used~\cite{markovicDevelopment}, as opposed to 
on-screen educational applications that emerged only in the last few decades~\cite{blume2011google, halle2017open, preim2018survey}.
Nowadays, biological and anatomical physicalizations are mainly restricted to model kits or contextual 3D printed models.
\rv{Although consumer-oriented 3D printers are becoming more and more popular, they can use a limited amount of colors and their maintenance remains costly and technically demanded for private usage~\cite{3Dprintingbasedonimagingdata}.}
Recently, a series of works proposed computer-aided pipelines for the easy generation of data sculptures, using affordable and readily available materials~\cite{Vol2velle, raidou2020slice, schindler2020anatomical, pahr2021}.  
The core motivation around such approaches relates to the notions of \rv{\textit{constructivism}~\cite{huang2010investigating} and \textit{embodiment}~\cite{jang2017direct}.
The former supports that active learning facilitates knowledge construction with less cognitive load, the latter claims that learning can benefit from the involvement of motorics and physical interaction.}

Among all possible types of data sculptures, \textit{papercrafts} are popular both as an art or recreational form~\cite{paczkowski2018papercraft3d} and as a tool serving educational purposes~\cite{eisenberg1998shop}.
Papercrafts are 2D or 3D objects from paper or cardboard that have been created by a 3D mesh unfolding into 2D patches, which can be printed and reconstructed back to 3D. 
Papercraft generation has proven to be a difficult task, given the often occurring overlaps and distortions~\cite{haenselmann2012optimal}, as well as the complexity of the reconstruction process~\cite{takahashi2011optimized}.
\chan{A discussion with a professional papercraft designer~\cite{Chan:2022} revealed to us that even ``simple'' models (i.e., with less than 100 polygon faces) require approximately a week for their creation and around ten days for the papercraft generation.}

\rv{Creating a papercraft with nested structures would add further to this complexity.
\textit{Nested papercrafts} would, for example, be needed for the investigation of a biological or anatomical model, which often include multiple, nested structures.} 
\rv{For example, a plant cell includes a nucleolus, which is nested within the nucleus. 
Subsequently, the nucleus is nested within the cytoplasm together with other structures, such as mitochondria, ribosomes, chloroplasts and the vacuole.}
Representing all these structures within one papercraft and ensuring their visibility, while maintaining the necessary level of interactivity and engagement without complicating the unfolding or the reconstruction, poses significant \rv{fabrication} challenges. 

In this work, we are investigating a new workflow for the computer-aided generation of \rv{nested} papercraft physicalizations that represent models from anatomy and biology. 
\rv{Figure~\ref{fig:teaser} shows an example of our nested papercrafts}.
Our proposed workflow targets specifically the representation of nested substructures within a 3D model and ensures their visibility.
In our approach, we employ hierarchical structuring to detect nested levels within a model, viewpoint calculation to ensure that outer levels can be cut for an unobstructed view on the inner levels, and a stability calculation at all levels (Figure~\ref{fig:teaser} (a--b)).
The final step is to unfold the nested levels of the model into 2D patches (Figure~\ref{fig:teaser} (c)), which can be reconstructed and examined (Figure~\ref{fig:teaser} (d)). 
As an additional visibility mechanism, we propose a smart strategy that takes advantage of the optical properties of our physical world by unveiling different channels of information through the use of different color channels (Figure~\ref{fig:teaser} (e)), inspired by our previous work~\cite{schindler2020anatomical}.
Our approach supports 3D models of varying nesting complexity, by exploiting the interactivity of nested papercraft models, and the projection of additional structures on their surface. 

The overall \emph{contribution} of this work is the design and realization of a workflow for the computer-aided generation of nested papercrafts to represent complex models from anatomy and biology.  
\rv{Our \emph{goal} is to support the generation of nested papercrafts in an easy (to manufacture and reconstruct), accessible (in resources and technologies) and affordable way. 
Our papercrafts can be used for tangible, interactive edutainment, where the user is engaged in the creation and assembly of the model's papercraft.}
The main \emph{components} of our approach include: 
\begin{itemize}
    \item The \textit{automated detection and representation of nested configurations} in anatomical and biological models, as well as corresponding papermesh synthesis.
    \item A dual (i.e., topology- and texture-based) approach to guarantee \textit{optimal visibility on the inner levels}.
    \item A strategy to ensure that the generated nested papercrafts are \textit{realizable} (i.e., assemblable and stable) and \textit{engaging}.
\end{itemize}

\vspace{-5pt}
\section{Related Work}
\label{sec:related}

\noindent\textbf{Anatomical and Biological Education and Edutainment:}
Preim and Saalfeld~\cite{preim2018survey} presented a survey on virtual anatomy education systems, where they review a big corpus of techniques targeting the education of medical students.
These span from surface and volume visualization~\cite{pommert2001creating, lichtenberg2016sline, vazquez2008interactive} to animations~\cite{bauer2014interactive}, and from anatomical labeling~\cite{ bruckner2005volumeshop} to virtual and augmented reality~\cite{saalfeld2016semi, pohlandt2019supporting, john2016use, Messier2016AnI3}.
Popular examples include VOXEL-MAN~\cite{pommert2001creating} and open anatomy browser~\cite{ halle2017open}.
Applications for the general public include the ZygoteBody~\cite{ blume2011google}.
For biological data, learning bio-molecular structures is achieved through gaming in VR~\cite{cai2006bio, cai2006immersive}. 
Other previous work focuses on teaching through hands-on graphics experiences that facilitate the exploration of structural aspects of macromolecular systems~\cite{canning2001teaching}.
Animation is also a fundamental concept that is often assessed to determine its suitability for educational purposes in biology~\cite{mnguni2021assessment}, as well as storytelling approaches~\cite{goodsell2021molecular}.
\rv{All the aforementioned approaches are on-screen solutions, which have proven to significantly contribute to anatomical and biological education, while also serving edutainment approaches either through gaming or storytelling. 
In our work, instead, we investigate how ``hands-on'' approaches involving physical models are applied to such scenarios.}

\noindent\textbf{Anatomical and Biological Physicalizations:}
Medical and biological data physicalizations are mainly 3D printed physical objects~\cite{Hybrid3Dprinting, 3Dprintingbasedonimagingdata, CardiacBloodFlowPhys}.  
Recently, Ang et al.~\cite{CardiacBloodFlowPhys} developed a cardiac blood flow physicalization based on 3D printing that allows the user to explore 4D MRI data in a slice-based manner.
On the biology side, Gillet et al.~\cite{gillet2004computer} combine AR and 3D printing in an application for structural biology education. 
Approaches that do not involve 3D printing are limited.
Historical approaches include the use of wax, wood, ivory, cardboard, and fabric models~\cite{markovicDevelopment}.
Computer-assisted approaches involve the use of volvelles, i.e., interactive wheel charts of concentric, rotating disks that support the physical fine-tuning of transfer functions~\cite{Vol2velle}, or sliceform papercrafts that support the representation of volumetric or mesh data using an octree-based partitioning~\cite{raidou2020slice}.
Both approaches target the visual and physical representation of the entire volume of a structure. 
However, the former focuses more on transfer function definition. 
The latter provides a holistic slice-based view of the data, but for higher resolution, more slices are needed, which increases the complexity of the approach.  
Interaction is also not possible without compromising the stability of the sliceforms.
Pahr et al.~\cite{pahr2021} proposed an interactive slice-based alternative for the physicalization of medical data that resembles holograms of volumetric medical data.
\rv{In our work, we also focus on largely available materials (i.e., paper), as this design choice can support the generation of physical models at home without the need for sophisticated solutions. Still, we focus on providing an illustrative physical representation of the data. }

\noindent\textbf{Papercraft Generation:}
The computer-assisted generation of papercrafts has been tackled before, for example for creating paper pop-ups from 3D meshes~\cite{ruiz2014multi, xiao2018computational} or origami architecture papercrafts~\cite{le2013surface}, iris papercrafts~\cite{igarashi2016computational},  \rv{cardboard-based papercrafts~\cite{zhang2016cardboardizer}}, and animated pop-ups that show motions of articulated characters~\cite{ruiz2015generating}.
This has also been investigated for large-scale models~\cite{sass2016embodied, chen2017generative} and entire 3D scenes through developable surfaces~\cite{paczkowski2018papercraft3d}.
Mesh unfolding has different applications, such as papercraft models~\cite{takahashi2011optimized,straubcreating} and models from self-folding materials~\cite{felton2013self,an2018thermorph}. 
Some techniques employ mesh deformation~\cite{chang2017improved,mitani2004making} to relax the unfolding problem.
The simultaneous handling of unfolding and glue tabs has been explored in our previous work~\cite{Korpitsch:2020:WSCG}.
An optimization model is formulated to search for an unfolding solution with optimal number of glue tabs needed for papercraft reconstruction.
However, to the best of our knowledge, none of \rv{the} aforementioned approaches research generating physical models that contain nested structures or the stability of such composite papercrafts.
This problem is not trivial, as it involves the ability to unfold individual meshes and the composition of the submeshes within a model.
\rv{In our work, we investigate thoroughly this aspect, applying it in the context of anatomical and biological edutainment.}

\vspace{-5pt}
\section{Requirements and Conceptual Choices}
\label{sec:overview}

\begin{figure*}[t]
    \centering{
    \setlength{\tabcolsep}{0pt}
    \begin{tabular}{cccccccc}
        \includegraphics[width=0.125\linewidth]{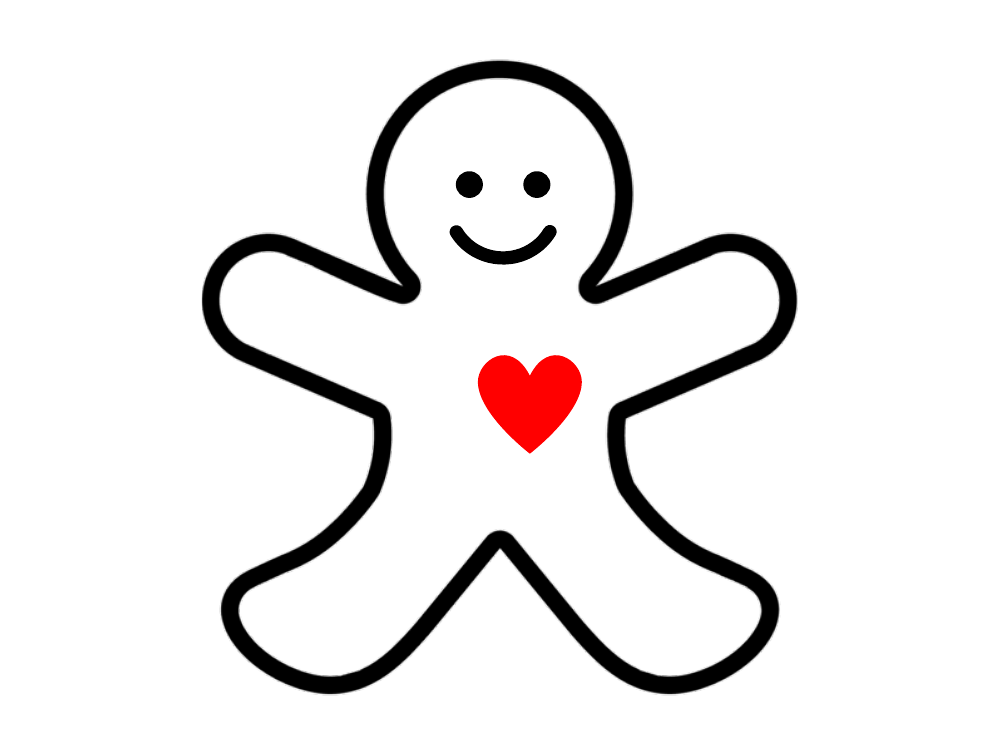} &
        \includegraphics[width=0.125\linewidth]{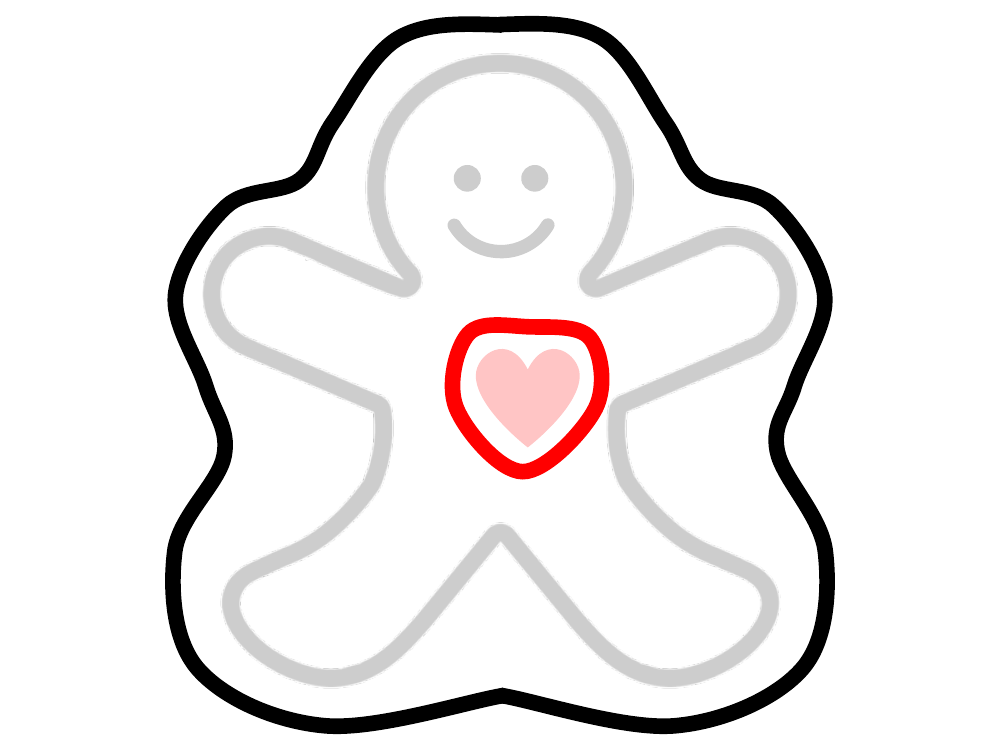} &
        \includegraphics[width=0.125\linewidth]{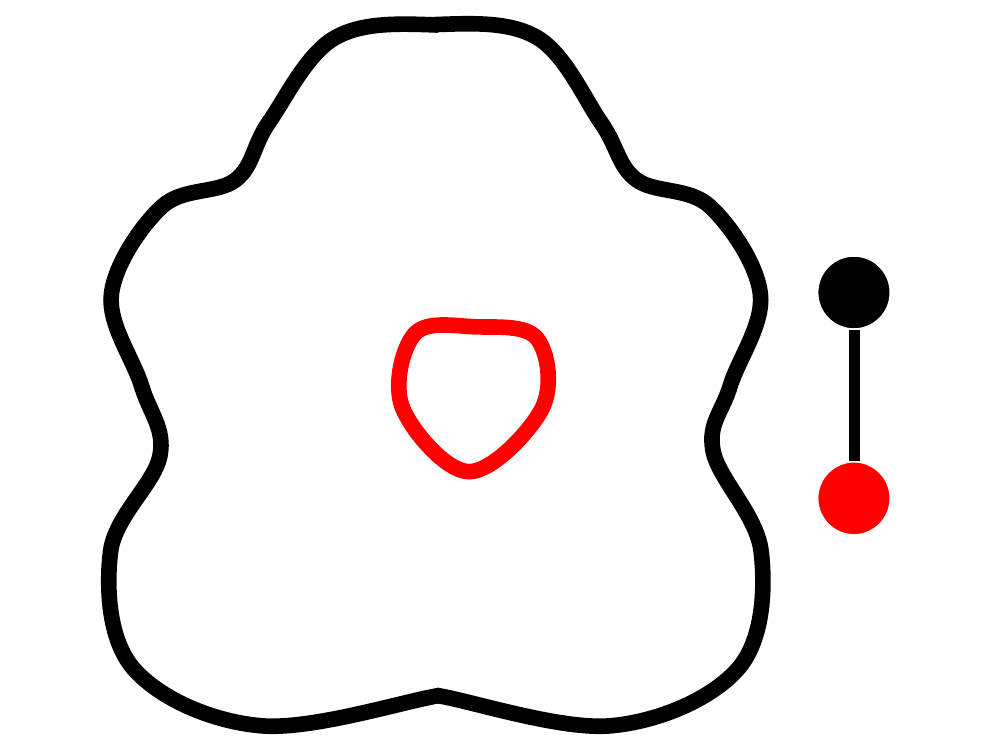} &
        \includegraphics[width=0.125\linewidth]{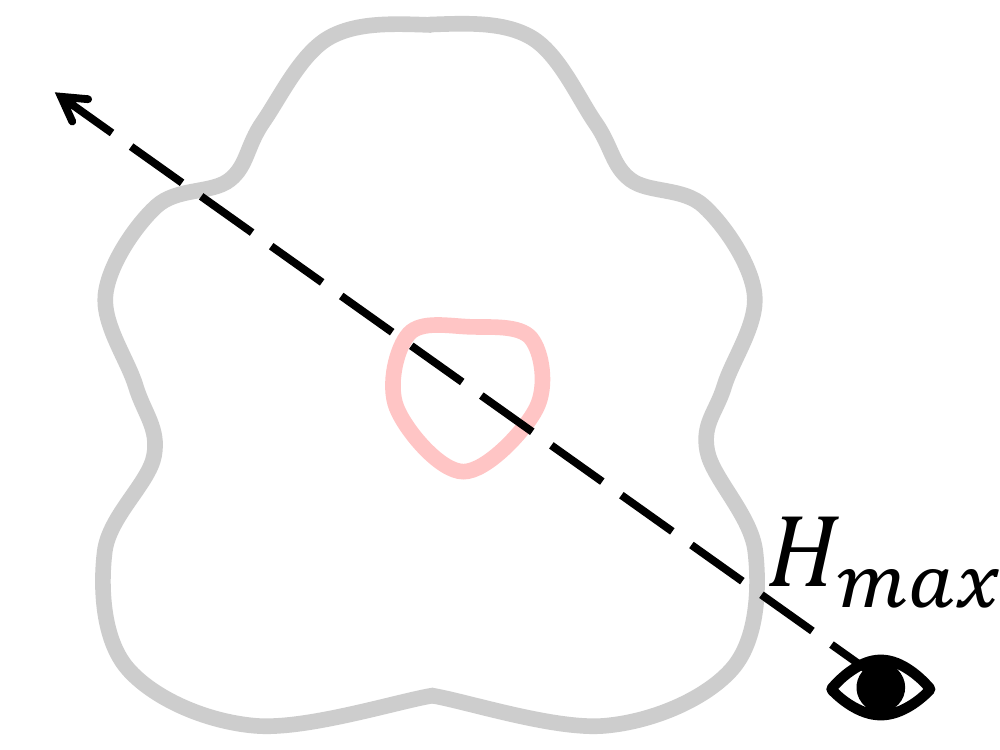} &
        \includegraphics[width=0.125\linewidth]{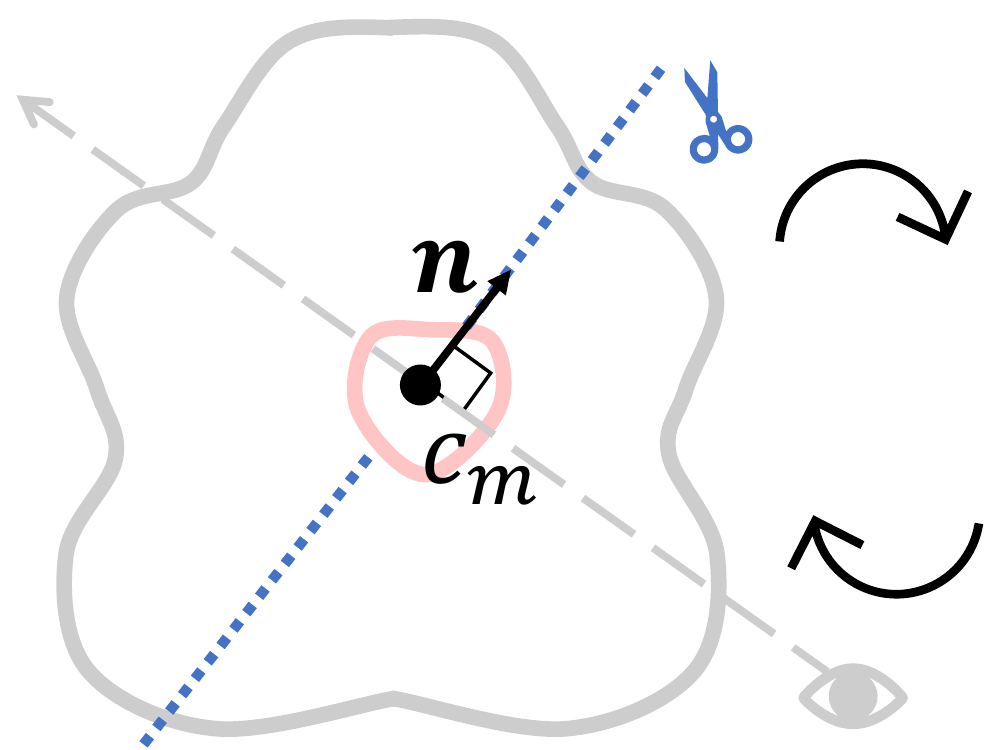} &
        \includegraphics[width=0.125\linewidth]{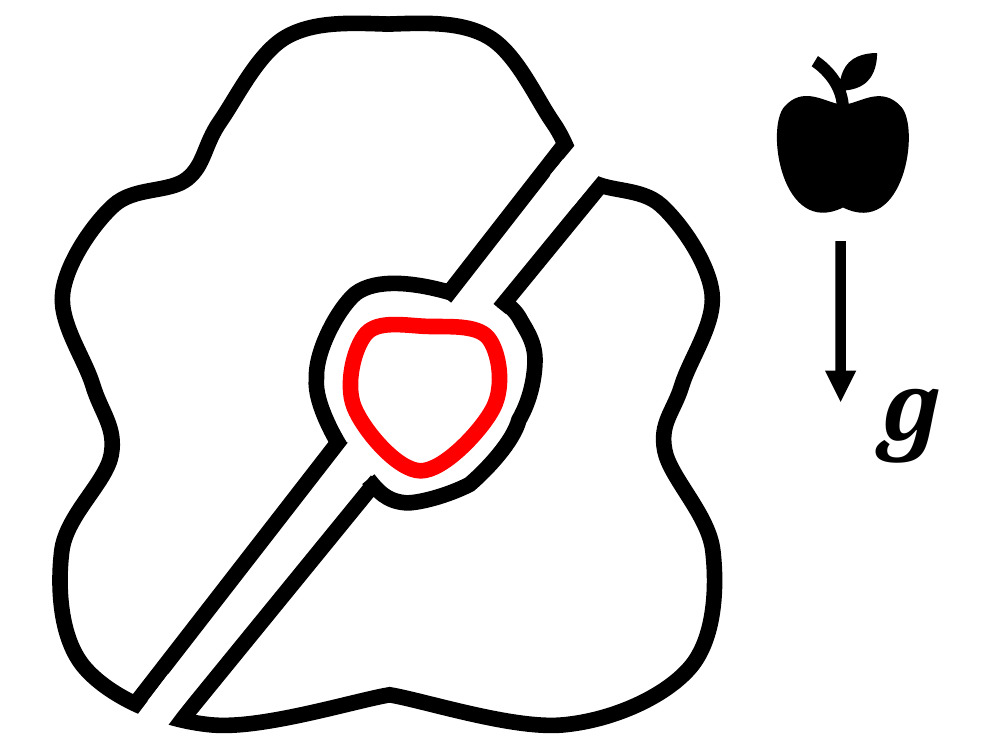} &
        \includegraphics[width=0.125\linewidth]{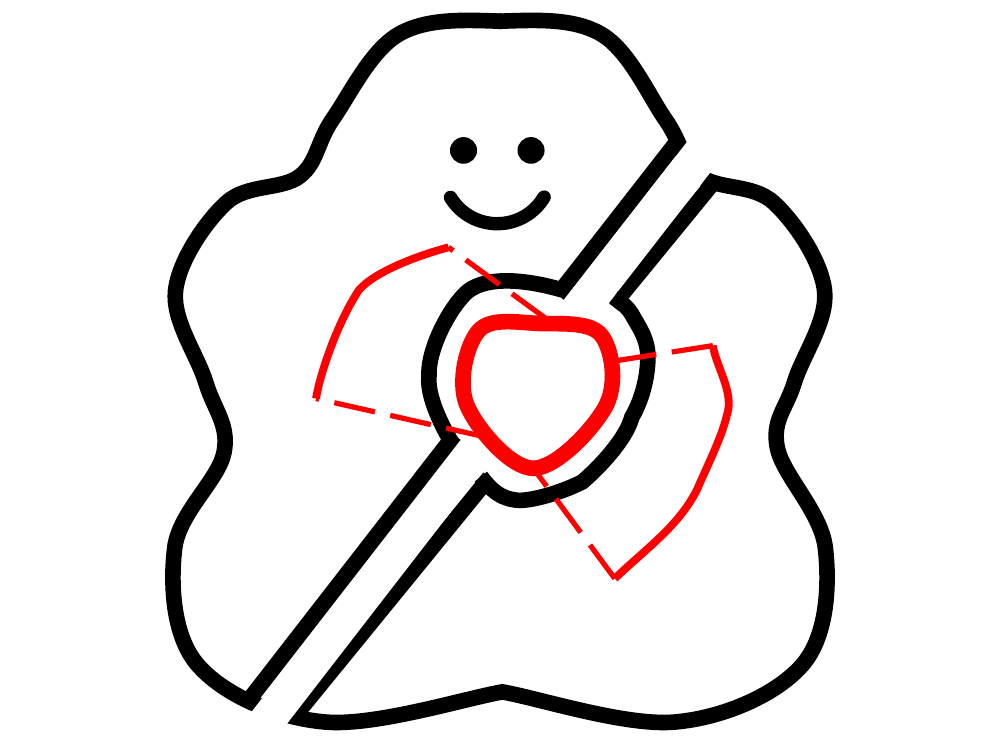} &
        \includegraphics[width=0.125\linewidth]{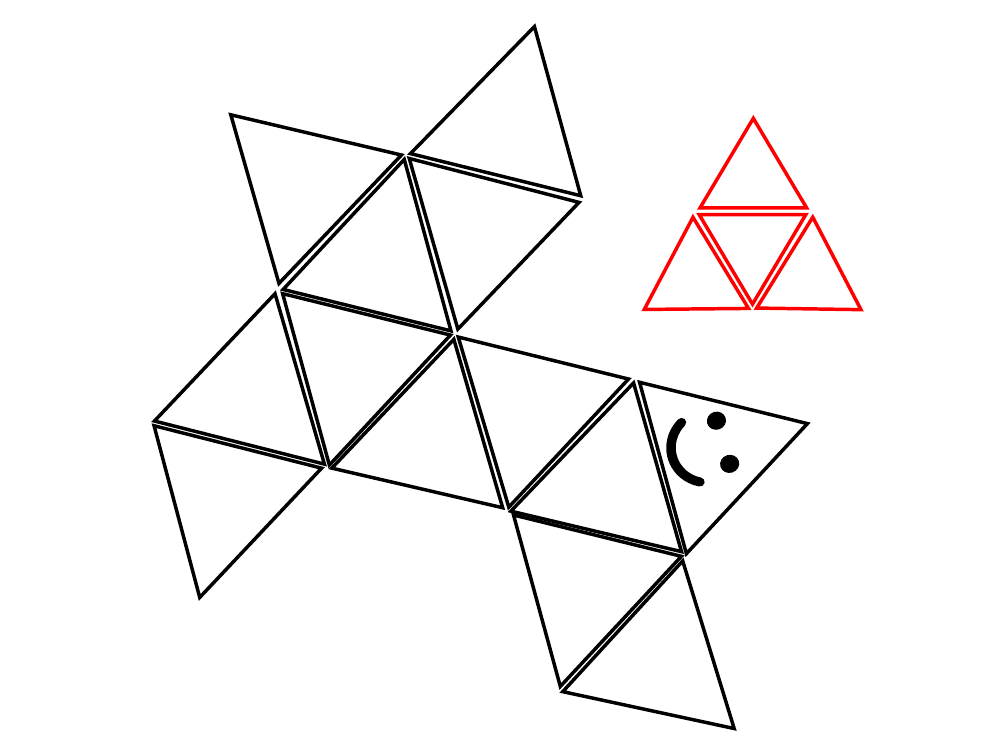} \\
        (a) Input  & 
        (b) Approximation & 
        (c) Hierarchy &
        (d) Best viewpoint & 
        (e) Cutting & 
        (f) Stability &
        (g) Projection &
        (h) Unfolding \\
    \end{tabular}
    \caption{Conceptual diagram depicting our workflow. We start by (a) taking as input a set of 3D meshes and (b) simplifying each mesh to create a coarse approximation. (c) We create a hierarchical structure that represents the nested configuration and (d) we perform an entropy-based viewpoint calculation to obtain the optimal view on each inner level, which is used to (e) cut the model open. \rv{(f) Stability calculations ensure that the constructed model does not fall apart and stands, and (g) our projection method renders textures on the surface of each level. Finally, (h)} we unfold each 3D model into a single 2D patch. \vspace{-5pt}}
    \label{fig:overview}
    }
\end{figure*}

The basic \textit{concept} of our approach is to provide a workflow for the cost-effective and easy generation of physical models that support nested structures, which are often present in biological and anatomical models. 
Building and exploring physical models aims to make the experience more engaging and, thus, more memorable or understandable~\cite{Embodiment}. 
Although papercrafts are not a new approach for education or edutainment~\cite{eisenberg1998shop}, the use of papercrafts to represent nested structures is novel and challenging.
\noindent Our main \textit{requirements} are: 
\begin{enumerate}
\setlength{\itemindent}{12pt}
    \item[\textbf{R1.}] \rv{The user does not need to have any knowledge of} the domain of application (i.e., anatomy or biology).
    \item[\textbf{R2.}] \rv{The approach requires a 3D mesh model as input} with multiple registered nested structures (submeshes). Examples could be a plant cell or an anatomical model of a human head.
    \item[\textbf{R3.}]  \rv{The user does not need to perform any complex interaction with our proposed pipeline (i.e., beyond few clicks) or to provide any other special input to it}.
    \item[\textbf{R4.}] The output of the workflow \rv{needs to} be an easy-to-assemble and engaging physical twin of the virtual model.
    \item[\textbf{R5.}] The assembly of the physical model \rv{needs to} be intuitive, and time- and cost-efficient.
    \item[\textbf{R6.}] The employed resources, i.e., materials and technologies, \rv{need to} be easily affordable and accessible. 
\end{enumerate}
 
\noindent Based on these requirements, our \textit{conceptual choices} are:
\begin{enumerate}
\setlength{\itemindent}{12pt}
    \item[\textbf{C1.}] \textbf{Supported data and structures:} 
    Our data are meshes representing anatomical and biological structures, which contain nested substructures, \rv{as dictated by \textbf{R2}}.
    These substructures are not ordered, i.e., there is no prior information about their topological hierarchy. 
    Also, we do not restrict the number of nested levels and we do not pose any limits to the number of children at each nested level.
    \item[\textbf{C2.}] \textbf{Nested papercrafts:}
    Our physicalizations are papercrafts with multiple nested levels, \rv{according to \textbf{R2}}. 
    The choice is made given the popularity of papercrafts, as well as the affordability and availability of materials \rv{(see \textbf{R6})}.
    \rv{We abstract the input meshes, so the papercraft can be assembled in a reasonable time.
    Given that the substructures are not ordered, we also detect the nested configuration and employ viewpoint selection to cut the outer levels and reveal the inner ones.} 
    This approach poses several challenges concerning the unfolding and reconstruction of the physicalization. 
    \rv{First of all, this step should not require the user to have prior knowledge of the domain \rv{(see \textbf{R1})}, nor to heavily interact with the framework to create the papercraft \rv{(see \textbf{R3})}.
    Furthermore, to} ensure that our papercraft is always feasible, i.e., can be unfolded and stable in reconstruction \rv{(see \textbf{R4})}, we abstract the structures of distinct levels to close-to-convex polygonal meshes. 
    \item[\textbf{C3.}] \textbf{Texture projections:}
    Conceptual choice \textbf{C2} does not account for three special cases: 
    \textbf{(a)} the mesh abstraction might introduce distortions, if the shape of the structure is inherently concave; 
    \textbf{(b)} the size of (inner) substructures might be prohibiting for printing and reconstruction, if these are too small; and 
    \textbf{(c)} structures with genus higher than $0$ or with complex topology, e.g., the surface of a Golgi apparatus might be problematic. 
    For these cases, we choose to use a smart visibility mechanism that projects textures from inner structures as additional information on the surfaces of the papercraft \rv{(see \textbf{R2})}. 
    This allows us to show correctly the shape of the inner structures on the approximated surfaces, even for small or intricate objects, while it can serve as ``\rv{X}-ray vision’’ to reveal the underlying levels. 
    Our projection is based on our previous work~\cite{schindler2020anatomical} and reveals up to three different inner structures with the use of colored filters. 
    \item[\textbf{C4.}] \textbf{Mesh unfolding into single connected patch:}
    Unfolding a 3D mesh into a single patch facilitates the reconstruction, as opposed to using multiple patches, but dealing with overlaps and distortions is a difficult task~\cite{takahashi2011optimized}. 
    In our previous work~\cite{Korpitsch:2020:WSCG}, we proposed a method for unfolding a 3D mesh into a single connected 2D patch by searching the best unfolding variant, which is represented by a minimum spanning tree (MST). A simulated annealing optimization is used to find an optimal unfolding. 
    The additional introduction of glue tabs supports the reconstruction process. 
    We make use of this approach, and we adapt it to accommodate choices \textbf{C2} and \textbf{C3}. \rv{This choice relates to \textbf{R5}}.
    \item[\textbf{C5.}] \textbf{Required materials and technologies:} 
    The reason for selecting papercrafts, as opposed to other digital fabrication methods, is that they are easy to manufacture \rv{(\textbf{R4})} and reconstruct \rv{(\textbf{R5})}, as they use affordable resources (paper and colored filters) and technologies (\rv{full-color 2D printers}), \rv{according to \textbf{R6}}.
\end{enumerate}

\vspace{-5pt}

\section{Nested Papercrafts Pipeline}
\label{sec:method}

\subsection{\rv{Pipeline Overview}}
\label{sec:overview_pip}
Figure~\ref{fig:overview} depicts the workflow of our approach. 
Our approach takes as input a set of \rv{registered} meshes that composes the 3D model of an anatomical or biological structure (Figure~\ref{fig:overview}(a), \textbf{C1}). 
In a preprocessing step, we simplify each mesh to create a feasible papermesh, i.e., a coarse approximation, as a basis for the approach (Figure~\ref{fig:overview}(b), Section~\ref{ssec:input}, \textbf{C2}).
In the next step, we use an iterative strategy to detect the nested levels within the model and to create a hierarchical structure that represents the nested configuration in the data  (Figure~\ref{fig:overview}(c), Section~\ref{ssec:hierarchical}, \textbf{C2}).
After obtaining the distinct nested levels, we perform an entropy-based viewpoint calculation step~\cite{vazquez2001viewpoint}, which returns the optimal viewpoint that provides an unobstructed view of each inner level (Figure~\ref{fig:overview}(d), Section~\ref{ssec:viewpoint}, \textbf{C2}).
Then, \rv{we cut each parent level} based on the optimal viewpoint on the child level, to reveal at opening time the best possible view on the inner level (Figure~\ref{fig:overview}(e), Section~\ref{ssec:cutting}, \textbf{C2}).
A stability estimation ensures that the cutting configuration at each level results in a feasible and stable physical model, i.e., the individual levels do not fall apart (Figure~\ref{fig:overview}(f), Section~\ref{ssec:stability}, \textbf{C2}).
Subsequently, a projection method creates textures to be rendered on the surface of the papermesh that represent a detailed version of the structure and/or additional inner levels or smaller structures, such as vessels (Figure~\ref{fig:overview}(g), Section~\ref{ssec:projection}, \textbf{C3}). 
For this, we take advantage of the subtractive color printing properties of light~\cite{schindler2020anatomical}, i.e., \rv{we use cyan, magenta, and yellow inks} to denote individual structures. 
\rv{We can reveal individual structures on demand} under appropriate colored filters (red, magenta, and blue, respectively). 
The final step is to unfold the multiple nesting levels of the 3D model with the additional texture projections into multiple single 2D patches~\cite{Korpitsch:2020:WSCG} (\textbf{C4}), which \rv{we can later print and reconstruct} (Figure~\ref{fig:overview}(h), Section~\ref{ssec:unfolding}, \textbf{C5}). 
Our approach exploits the interactivity of nested papercrafts, as well as the projection of additional structures on their surface to allow the examination of 3D models of varying complexity.
\vspace{-5pt}

\subsection{Input and Data Approximation}
\label{ssec:input}

Our input data are \rv{registered} meshes from anatomical and biological models with a nested configuration \textbf{R2}. 
No other prior information is required about them, as discussed in Section~\ref{sec:overview}, \textbf{C1}.
Also, as discussed in \textbf{C2} and \textbf{C3}, our fundamental idea is to use an approximation of the input mesh to create a papercraft model that represents the nested configuration, while \rv{we project} additional structures as textures on the surfaces of the papercraft. 
\rv{In our previous work we already proved that projection can be an effective strategy}~\cite{schindler2020anatomical}, \rv{while we also demonstrated the need for following more organic shapes for the crafts~\cite{raidou2020slice}---even if these are simplified}.
We need two types of minimal user input (\textbf{R1,3}) for our papercraft physicalization: \rv{which meshes to group together onto a papercraft and which color to use for each.}
In our current implementation, users choose which structures are particularly interesting for them to use in the projections.


In this paper, we use the terms ``\textit{papercraft}'' to denote the final output of the physicalization workflow and ``\textit{papermesh}'' to denote the virtual twin of the papercraft.
\rv{Each of our papermeshes is a manifold triangular mesh with genus $0$~\cite{munkres2000topology}.}
Before we can create our nested papercraft, we need to approximate the fine input mesh to a reasonably foldable papermesh \textbf{(R4,5)}. 
Takahashi et al.~\cite{takahashi2011optimized} showed that the participants could construct meshes with around $150$ faces, while meshes over $300$ faces were less feasible. 
Although, theoretically, we support papermeshes with a high face number, 
we follow Takahashi et al. and limit the total face number to $150$.
\rv{The upcoming sections provide details on the achieved sizes for several anatomical and biological models.} 

\begin{figure}[b]
    \centering{
    \includegraphics[width=.8\linewidth]{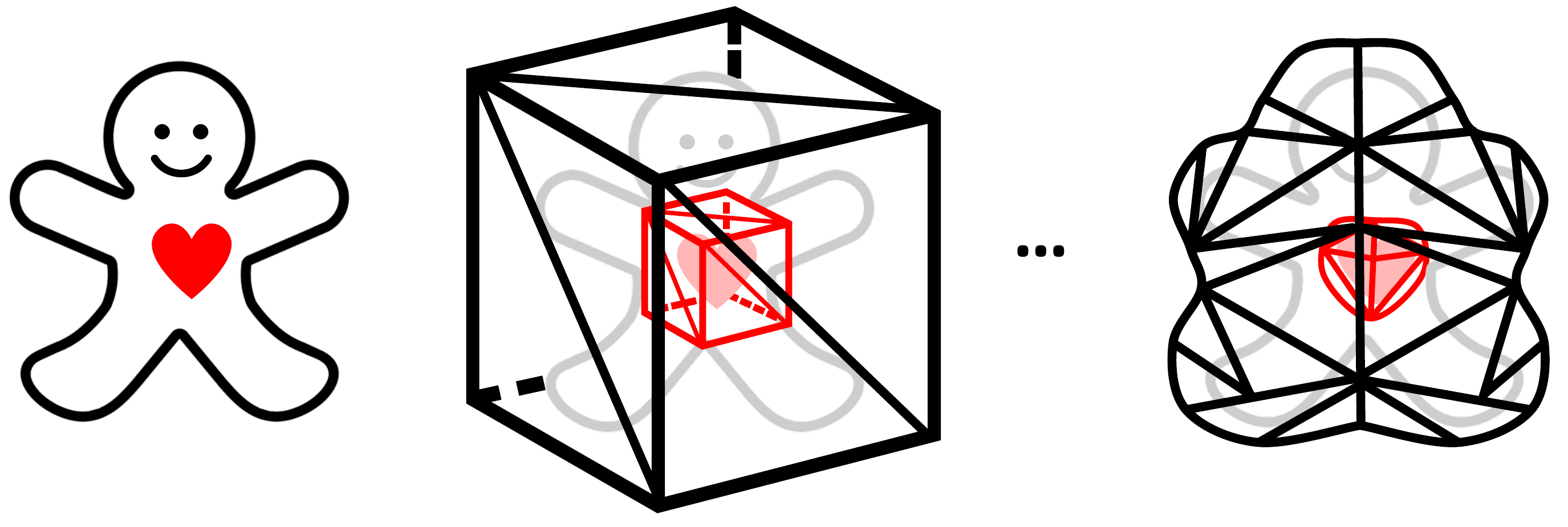}
    \caption{Conversion of input structures to a set of papermeshes.}
    \label{fig:input}
    }
\end{figure}

In this preprocessing step, \rv{we create our papermeshes} in an iterative manner. 
We start by initializing a bounding box that wraps each loaded structure, as shown in Figure~\ref{fig:input}. 
Subsequently, we subdivide the bounding box into a fine triangular mesh
\rv{(e.g., $3,000$ faces for a large anatomical model such as the head)}, 
and we move each vertex of the bounding box to the corresponding closest vertex on the structure. 
This process allows us to create a compact wrapper that retains better the shape of the structure. 
Once we have such an \rv{approximating} mesh, we apply the mesh simplification algorithm by Lindstrom and Turk~\cite{cgal:sm-simplification} to significantly reduce the mesh face number and to create a papermesh that \rv{we can unfold later}.

\vspace{-5pt}

\subsection{Hierarchy Detection}
\label{ssec:hierarchical}

\begin{figure}[b]
    \setlength{\tabcolsep}{10pt}
    \centering{
    \includegraphics[width=.8\linewidth]{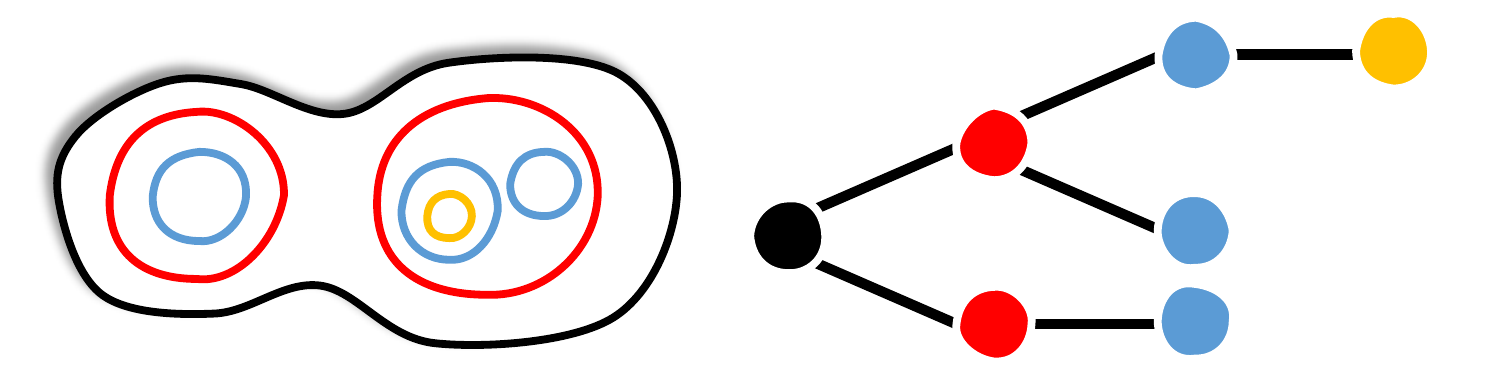} \\
    \caption{Conceptual diagram of a hierarchical relationship in a nested model using a tree data structure.}
    \label{fig:tree}
    }
\end{figure}

Once \rv{we create the papermesh approximations}, we need to detect the hierarchy within the nested configuration of the entire model to ensure we can open up outer levels to show inner ones \textbf{(C2, R2)}.
We identify meshes that are encapsulated in others, and we use this information to build a tree $T$ representing this relationship. Figure~\ref{fig:tree} shows an example of a tree constructed for a given nested model.
\rv{We solve this by using a tree insertion algorithm.} 
As input we consider an unsorted set of papermeshes $M = {m_a, m_b, ..., m_n}$. 
The first mesh $m_a$ is the root of $T$. 
\rv{We need to check all following meshes} for their hierarchical relationship to the first one. 
Consider the second mesh to insert $m_b$. 
\rv{If $m_a$ encapsulates $m_b$, we assign $m_b$ as a child node of node $m_a$ in $T$. 
Otherwise, if $m_b$ encapsulates $m_a$, we assign $m_b$ as the new root and $m_a$ as a child of $m_b$. 
We continue this process and traverse $T$ (Breadth-First Search, BFS or Depth-First Search, DFS) to find the right position of each papermesh.}

\rv{We conduct the \textit{mesh-inside test} in two steps.}
First, we test if all vertices on $m_b$ are inside $m_a$~\cite{cgal:atw-aabb-21b}, and then we perform dD Iso-oriented box intersection tests~\cite{cgal:lty-pmp-21b} on the edges to detect edge--edge intersections. 
If the first step is valid and \rv{we find no intersection, we settle the relationship between $m_a$ and $m_b$.}
Note that the algorithm runs under the assumption that no papermeshes intersect, as this is the case for anatomical and biological structures.


\vspace{-5pt}

\subsection{Viewpoint Calculation} 
\label{ssec:viewpoint}

\begin{figure}[b]
    \centering{
    \includegraphics[width=\linewidth]{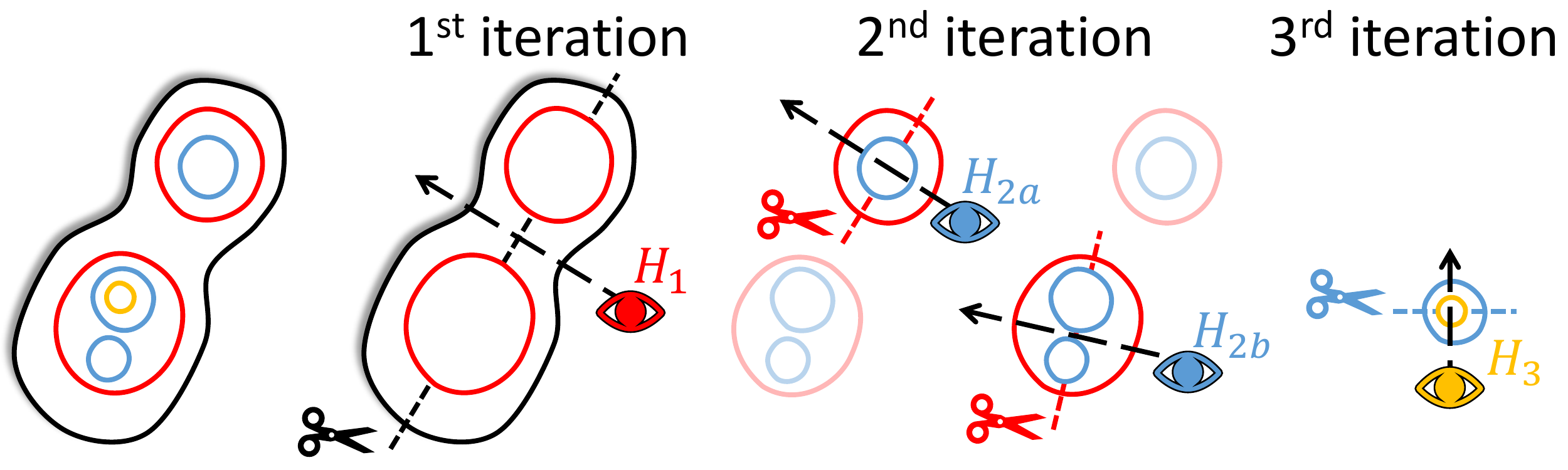} 
    \caption{Example of the viewpoint calculation within the nested structure. The viewpoint at level $l$, where $l=1,2,3$ with maximum entropy $H_l$ is used to cut the parent level $l-1$. }
    \label{fig:viewpoint}
    }
\end{figure}

After capturing the hierarchy within the nested structure, we need to find a suitable configuration to cut it open \textbf{(C2, R2)}. 
A suitable cutting plane must fulfill two conditions.
First, when opening an outer nested level, we need to provide the \textit{best view on the inner levels}. 
We define the best view, as a view that reveals the inner structure in the most recognizable manner, as soon as the outer structure is open.
Second, cutting the outer structure should \textit{not affect the stability} of the entire nested model. 
Although the best view needs to take the stability into account, we deal with this further together with the next two steps of our workflow (Sections~\ref{ssec:cutting}--~\ref{ssec:stability}). 

Optimal viewpoint calculation is a heavily researched topic, relevant for several applications within computer graphics and visualization.
We refer \rv{our readers} to a recent comparative survey on viewpoint calculation methods by Bonaventura et al.~\cite{bonaventura2018survey}. 
Our definition of the best view within our specific application is the view that supports the recognizability of the inner levels when opening the outer level.  
Reviewing existing literature~\cite{bonaventura2018survey, dutagaci2010benchmark, secord2011perceptual, polonsky2005s} points to entropy-based methods, such as the maximum viewpoint entropy calculation by Vazquez et al.~\cite{vazquez2001viewpoint}.
This robust measure optimizes in a balanced way the projected area and the number of faces that are visible.
Measures that target only area (or silhouette) optimization would not be sufficient, as these measures do not tell us about the amount of detail we can see in our scene. 
Measures that target only the number of faces are also insufficient, as scenes with a high number of small faces would be inadequately described.
The maximum viewpoint entropy indicates that a certain viewpoint balances the number of visible faces with respect to the relative projected area.
Although Sbert et al.~\cite{sbert2005viewpoint} found that this method is sensitive to discretization, this is not an issue for our approach, while we discarded other approaches~\cite{viola2006importance} due to high complexity.

Figure~\ref{fig:viewpoint} shows the steps of our viewpoint calculation. 
We conduct a recursive calculation based on viewpoint entropy, as defined by Vazquez et al.~\cite{vazquez2001viewpoint} at each level of our nested model (except for the outermost level $l=0$). 
The viewpoint entropy $H$ at each level $l\geq1$ assumes as probability distribution the relative area of the projected faces over a sphere of directions centered in the viewpoint, and is defined as: $H_{l} = -\sum_{i=0}^{^{N_l}}\frac{A_{i,l}}{A_{t}}log\frac{A_{i,l}}{A_{t}}$, where $N_l$ is the number of faces of the scene at a specific level $l$, $A_i,l$ is the projected area of face $i$ of level $l$ over the sphere, and $A_t$ is the total area of the sphere.
The maximum viewpoint entropy $H_l$ at each level $l$ is used as input to cut the outer level $l-1$ (Figure~\ref{fig:viewpoint}), to reveal the structure of the inner level. 
For example, \rv{we use} the viewpoint at level $1$ that gives the maximum entropy $H_1$ (Figure~\ref{fig:viewpoint}, see ``1st iteration'', red) to cut the parent level (black).
\rv{We cut the other levels similarly} (Figure~\ref{fig:viewpoint}).
\vspace{-5pt}

\subsection{Optimal Cutting Plane} 
\label{ssec:cutting}

\begin{figure}[b]
    \centering{
    \setlength{\tabcolsep}{0pt}
    \begin{tabular}{cc}
        \includegraphics[width=0.5\linewidth]{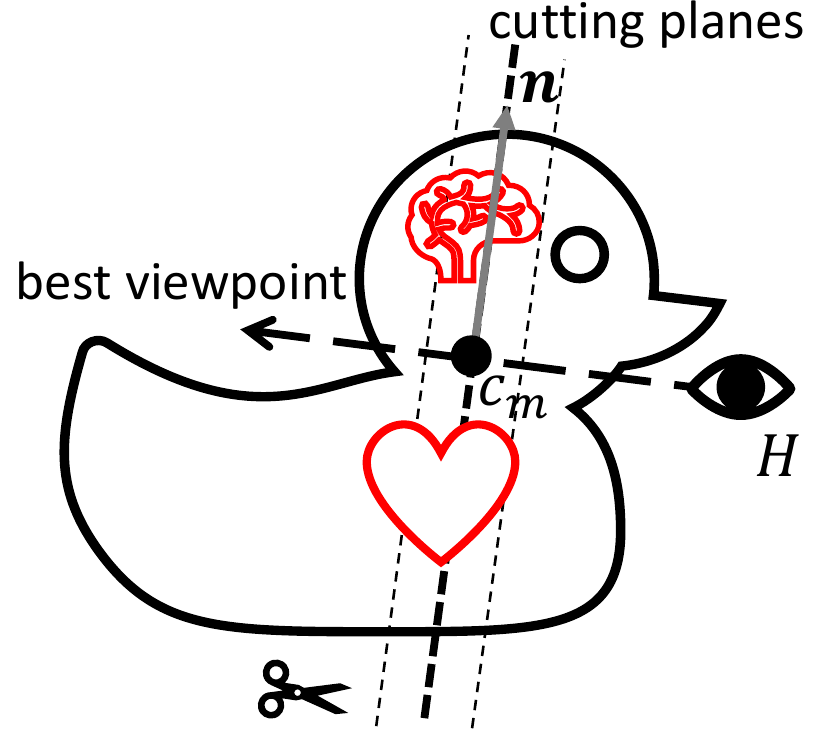} &
        \includegraphics[width=0.5\linewidth]{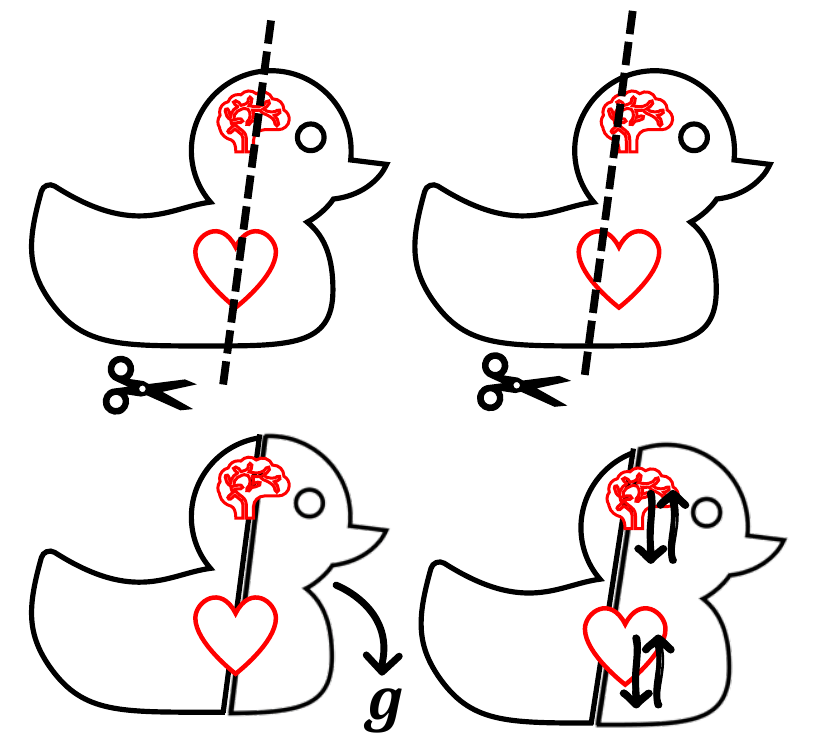} \\
        (a) Cutting plane & 
        (b) Stability testing \\
    \end{tabular}
    \caption{Example of (a) a cutting plane (perpendicular to the best viewpoint direction and passing through the center of mass $c_m$ of the inner level) and (b) testing for a stable papercraft composition.}
    \label{fig:cutting}
    }
\end{figure}

Figure~\ref{fig:cutting} shows the steps of the optimal cutting based on the viewpoint calculation. 
To provide the optimal cutting plane \textbf{(C2, R2)}, we need to conduct the following steps. 
First, we subtract the meshes of the inner level from the outer level to create an ``envelope'' mesh that carries the difference operation between the inner and outer level.
Then, we compute the center of mass $c_m$ of the meshes of the inner level and we place at this location the origin of our cutting plane. 
\rv{We retrieve the normal \textbf{$n$} from the viewpoint calculation}, as described in the previous section (Figure~\ref{fig:cutting} (a)). 
Subsequently, \rv{we clip the envelope mesh (between the outer and inner level)} using the cutting plane as an implicit function. 
This returns two meshes with two boundary edge loops, each.
\rv{We, then, stitch these together with triangles to ensure continuity.} 
\vspace{-5pt}

\subsection{Papercraft Stability Test}
\label{ssec:stability}

One important factor for a nested papercraft is the stability of the constructed model in the physical world \textbf{(C2, R2,4)}. 
The assembled nested papercraft should not fall apart, and the subparts should stay connected withing the entire physical model. 
We examine the papercraft stability using physical simulation.
Note that the stability cannot account for errors introduced by the users during crafting, which cannot be easily controlled as they strongly depend on the user's experience.
\rv{We conduct the stability test iteratively} in conjunction with the viewpoint and the cutting plane calculation (Sections~\ref{ssec:viewpoint}--~\ref{ssec:cutting}). 
Each cut should guarantee that the decomposed subparts will integrate well with each other.
Otherwise, an unstable cut will cause the model to fall apart (Figure~\ref{fig:cutting} (b), left column), while a slightly adapted cut ensures a stable composition based on gravitational and frictional forces (Figure~\ref{fig:cutting} (b), right column).
The articulated body algorithm by Featherstone~\cite{featherstone1987aba} for physics simulation is sufficient for our simple scenes. 
This is a forward dynamics approach with computational complexity $O(n)$. 
In our implementation, we use gravity as an external force, we assign a mass to each individual papermesh, and simulate if papermeshes will fall apart in a physical environment. 
If so, the algorithm will return an \emph{unstable} flag back to the cutting plane algorithm. 
The cutting plane algorithm will look for the next best plane until we reach stability.
\vspace{-5pt}

\subsection{Texture Projection}
\label{ssec:projection}

One papermesh can contain multiple substructures (e.g., an anatomical mesh contains bones, muscles and vessels) \textbf{(C2--3, R2)}.
To ensure the visibility of the different substructures, we use a strategy, adapted from our previous work~\cite{schindler2020anatomical}.
This strategy allows us to use up to three channels to render different substructures on the surface \textbf{(C3)}, which \rv{we can bring forward} through the use of red, green, and blue colored filters or light \textbf{(C5)}. 
These colored filters isolate (groups of) structures rendered respectively in cyan, magenta, and yellow, taking advantage of the properties of visible light.
\rv{We show an example of this in Figure~\ref{fig:teaser} (e).}
For more details on the color filter mechanism, we refer to our previous work. 

\begin{figure}[b]
    \centering{
    \setlength{\tabcolsep}{0pt}
    \begin{tabular}{ccc}
        \includegraphics[width=0.33\linewidth]{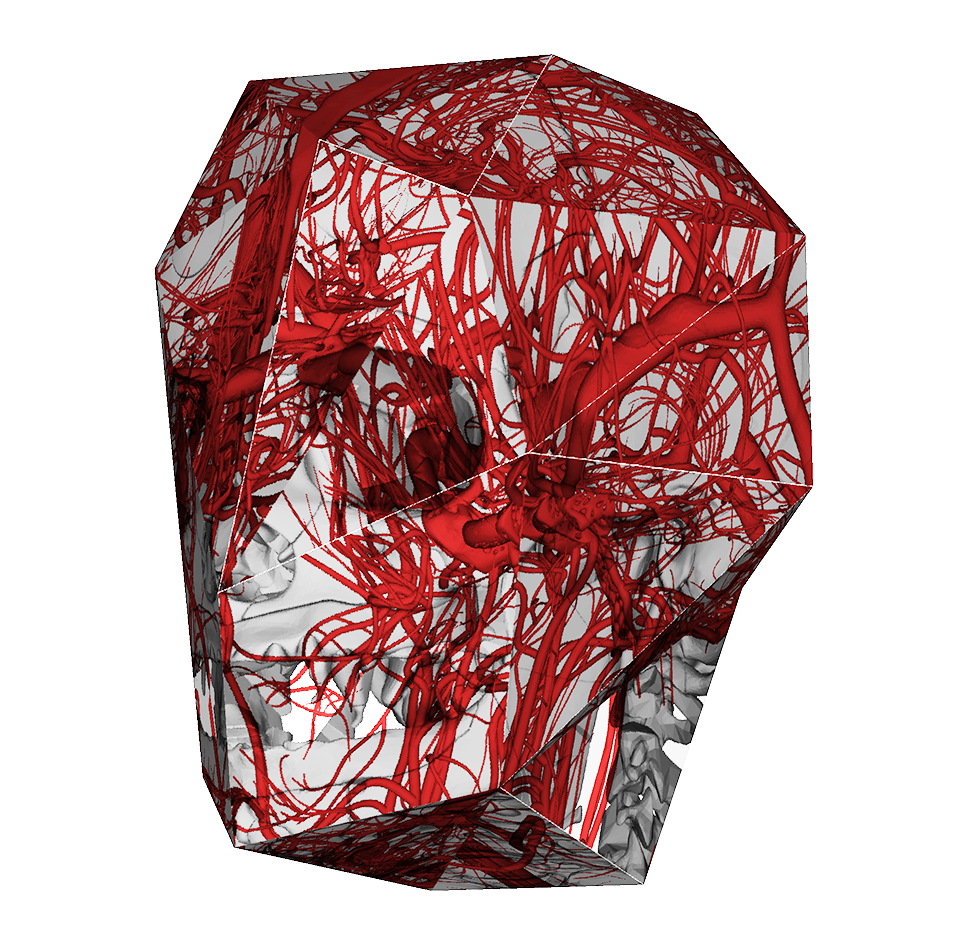} &
        \includegraphics[width=0.33\linewidth]{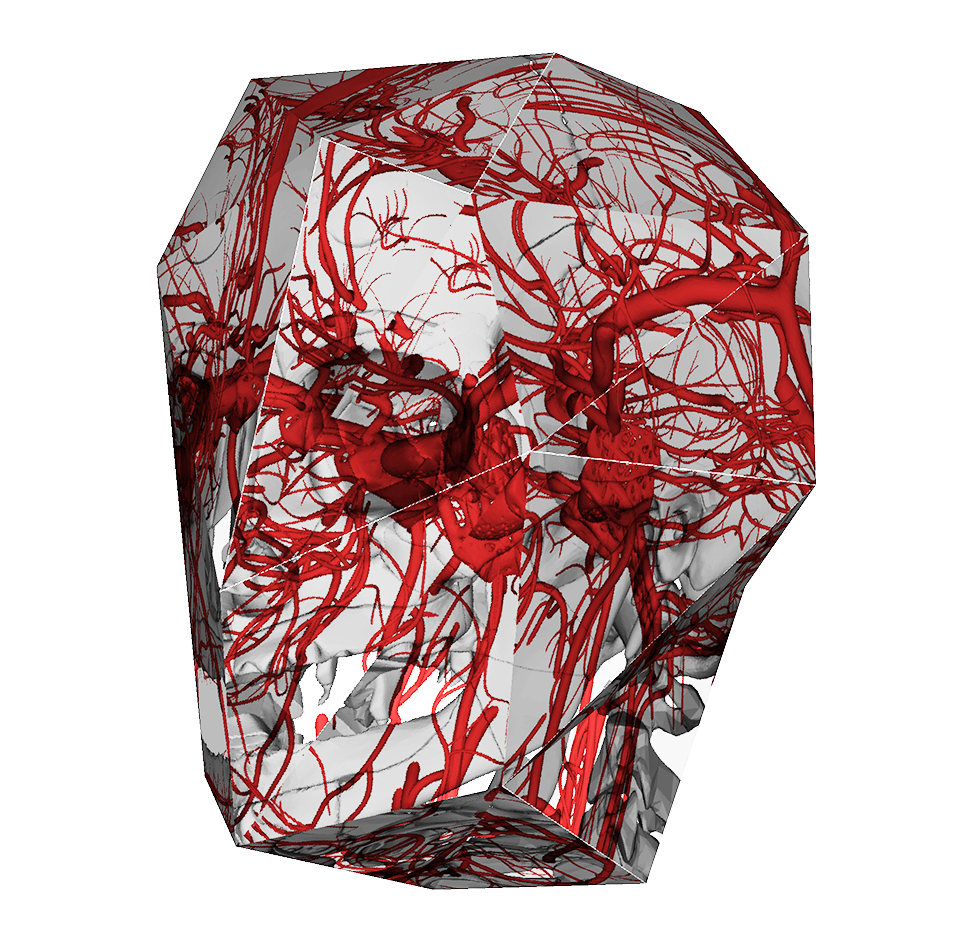} & 
        \includegraphics[width=0.33\linewidth]{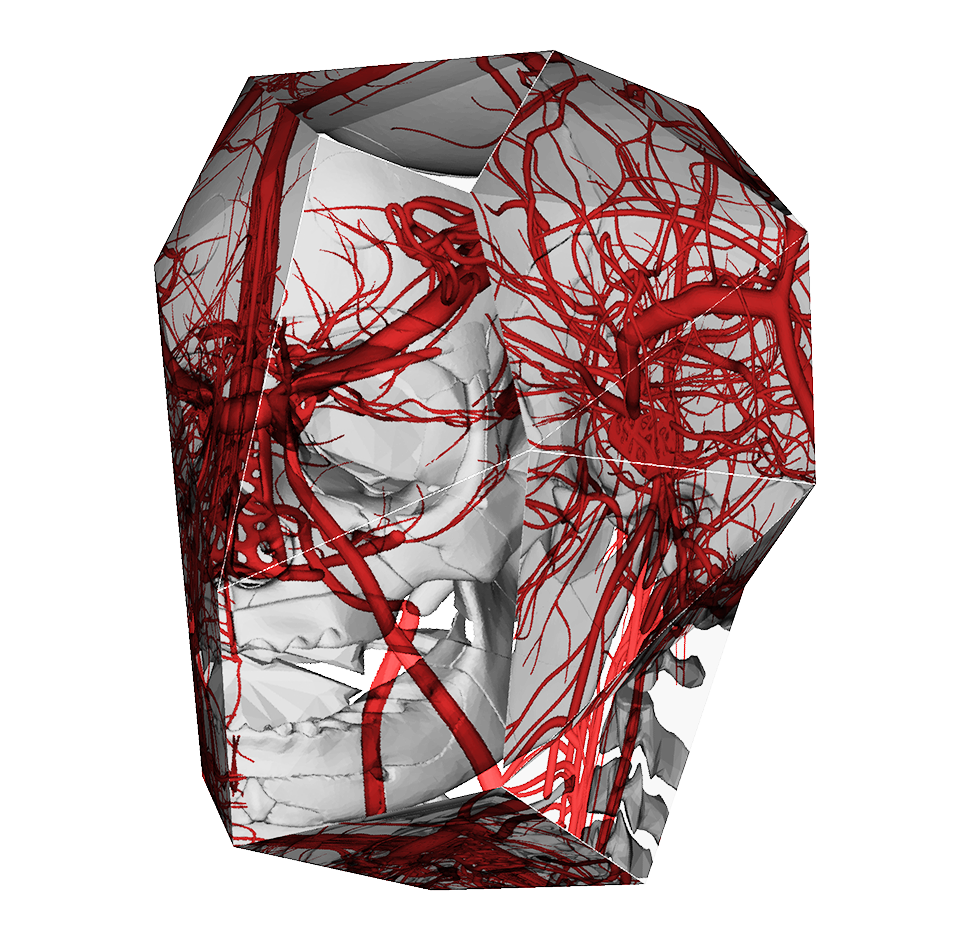} \\
        (a) & 
        (b) &
        (c) \\
    \end{tabular}
    \caption{Different modes of our projection technique~\cite{schindler2020anatomical}: (a) inflation, (b) clipping, and (c) cube mode.}
    \label{fig:proj_modes}
    }
\end{figure}

For each structure, we iterate over the triangles of the respective papermesh and we project on them a texture. 
Before starting the projection, the user can adapt the papermesh to fit better the individual structures using one of the following three modes: 
\textit{inflation mode} wraps the papermesh around a single structure, projects and then, inflates it back to the original; 
\textit{clipping mode} removes features that are more distant from the texture without changing the papermesh; 
and \textit{cube mode} projects the papermesh faces onto the faces of the bounding cube along the face normals of the cube. 
\rv{Figure~\ref{fig:proj_modes} summarizes these modes.}
For papermeshes that have multiple pieces, the bounding cube is also cut into multiple pieces, using the same cutting plane.
To construct the texture for each triangle, we place a camera at the center of the triangle facing towards the structure.
\rv{We map the values for each texture to a single hue} (magenta, cyan, or yellow) sequential colormap, where higher values are mapped to higher luminance, i.e., $[t_{min}, t_{max}] \rightarrow (h,s,v)=(200,240,120...240)$, as $(h,s,v)=(200,240,120)$ corresponds to pure magenta and $v=240$ to white for all values of $h$ and $s$.
Subsequently, \rv{we normalize each texture to the range $[0,1]$ and we multiply with the other textures of this papermesh, if available}.


\vspace{-5pt}
\subsection{Papermesh Unfolding}
\label{ssec:unfolding}

The last step of our workflow generates manifold papermeshes to be unfolded individually.
To follow up on \textbf{(C4, R4--5)}, we do not allow mesh deformation to retain the shape approximation, as described in Section~\ref{ssec:input}.
We also aim to keep the unfolded patches as single connected components, so that the reader gains an overview on the number of individual papercrafts.
To achieve this, we adapt our previous approach~\cite{Korpitsch:2020:WSCG}.

\rv{We handle the mesh unfolding problem using properties} from graph theory, i.e., we use a spanning tree to represent an unfolding patch~\cite{takahashi2011optimized}. 
\rv{We create the spanning tree by building a dual graph of the input papermesh.}
Then, we find the minimum spanning tree of this dual graph. 
\rv{We use the weight assigned to the edges in the spanning tree search algorithm to control the priority of a boundary edge to be cut. }
For example, if an edge $e$ is a sharp mountain or valley fold, we avoid cutting this edge since it increases the complexity during construction. 
In this case, the edge $e$ is likely to be added in the tree. 
For more details about the full criteria, we refer \rv{our readers} to our previous work~\cite{Korpitsch:2020:WSCG}.
Mathematically, \rv{we use the spanning tree to describe variations of unfolded patches. The tree also serves as a backbone to control the unfolding process from 3D to 2D.}
In addition, \rv{we introduce glue tabs} to facilitate stitching.
This adds another complexity to the optimization, as the triangles of the input mesh should not intersect with the glue tabs when unfolded. 
For this, we use a simulated annealing optimization.

In our previous work, we did not consider textures on the meshes and used only indicators to guide the users during reconstruction. 
In the current approach, we consider textures as rendered from the previous projection step, but we still need the reconstruction indicators. 
For this, we take the advantage of double-sided printing. 
On one side of the paper we render the glue tabs, and on the other side we render the corresponding position to be glued on the back side of the patch.
In this way, \rv{we can render additional indicators on the back of the paper} without interfering with the texture of the structure projected on the front side.
Figure~\ref{fig:result_real_1} (c-1) shows an example of the front side and (c-2) an example of the back side.

\vspace{-5pt}
\subsection{Nested Papercraft Construction and Interaction} 
\label{ssec:papercraft}

Once \rv{we generate the unfolded patches, we print out these} on A4 or A3 papers and follow the glue tab instructions to create the papercraft physicalization \textbf{(C5, R4--6)}.
The assembly requires double-sided tape or glue.
An advantage of unfolding all papermeshes to connected patches (one patch for each level of the papermesh) is to provide the users with an overview of the total number of papermeshes in this 3D model \textbf{(C4, R4--5)}.
Once \rv{we assemble the papercraft model, we can open up} the subcomponents and explore the information projected on the surface with the use of colored filters \textbf{(C5, R6)} to bring forward different structures (Figure~\ref{fig:teaser}(e)).
\vspace{-5pt}

\subsection{Implementation}
\label{ssec:implementation}
\rv{The main source code is in C++17 and Python 3.9. 
Major libraries include VTK 
for fundamental computer graphics management, PyQt5 
for the user interface, and CGAL 
for computational geometry. 
\rv{We compiled the code} on a Windows 10 machine using MSVC 
and CMake. 
\rv{We used Pybullet~\cite{coumans2021}} for the physical simulation, and OpenGL for rendering the final results to an off-screen frame buffer using a camera facing the center of the unfolding using parallel projection.
Our implementation, along with examples and a demo of this work, is available at our repository~\cite{repo}.
} 

\begin{figure}[b]
    \centering{
    \setlength{\tabcolsep}{0pt}
    \begin{tabular}{cc}
        \includegraphics[width=0.43\linewidth]{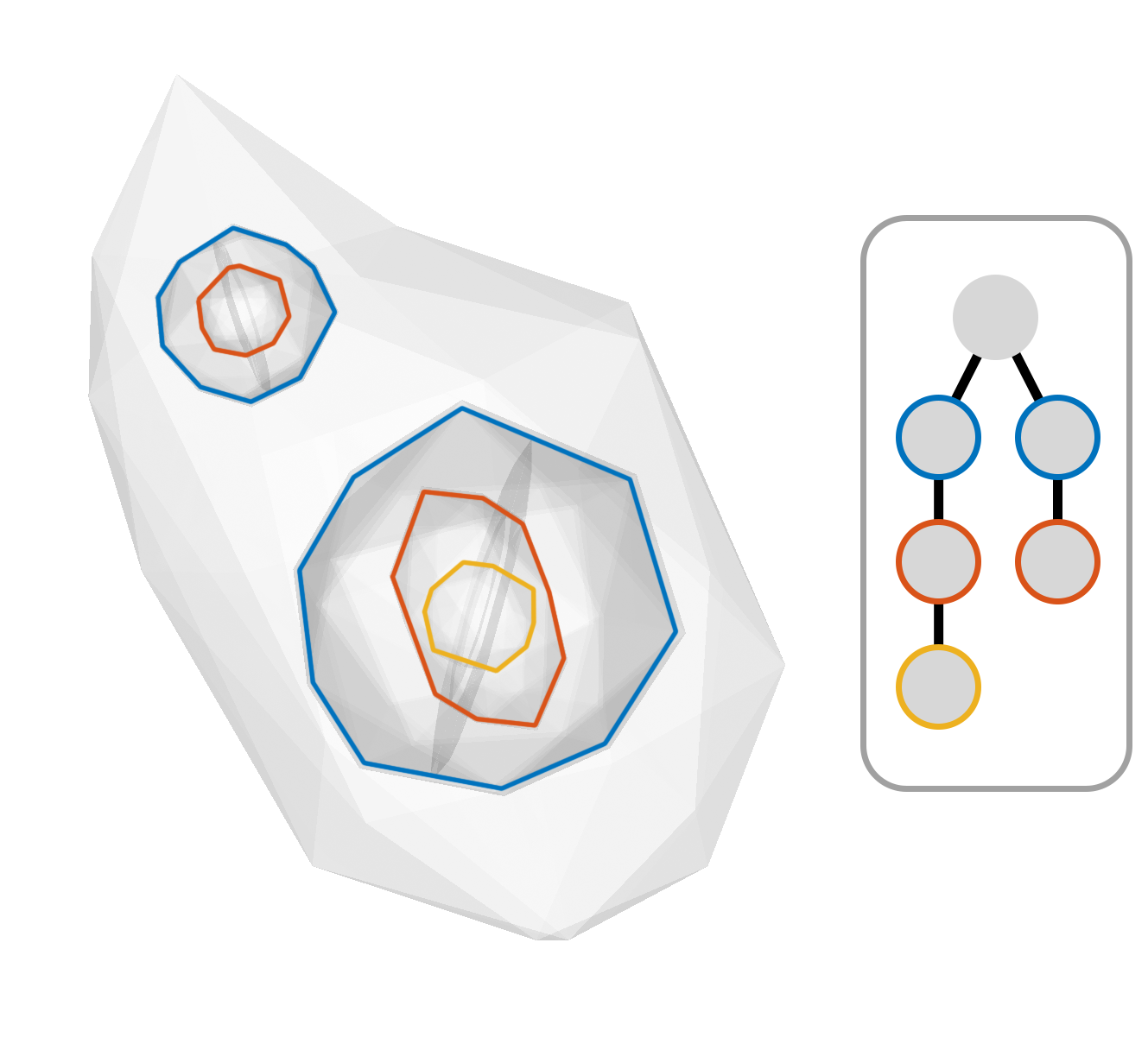} &
        \includegraphics[width=0.43\linewidth]{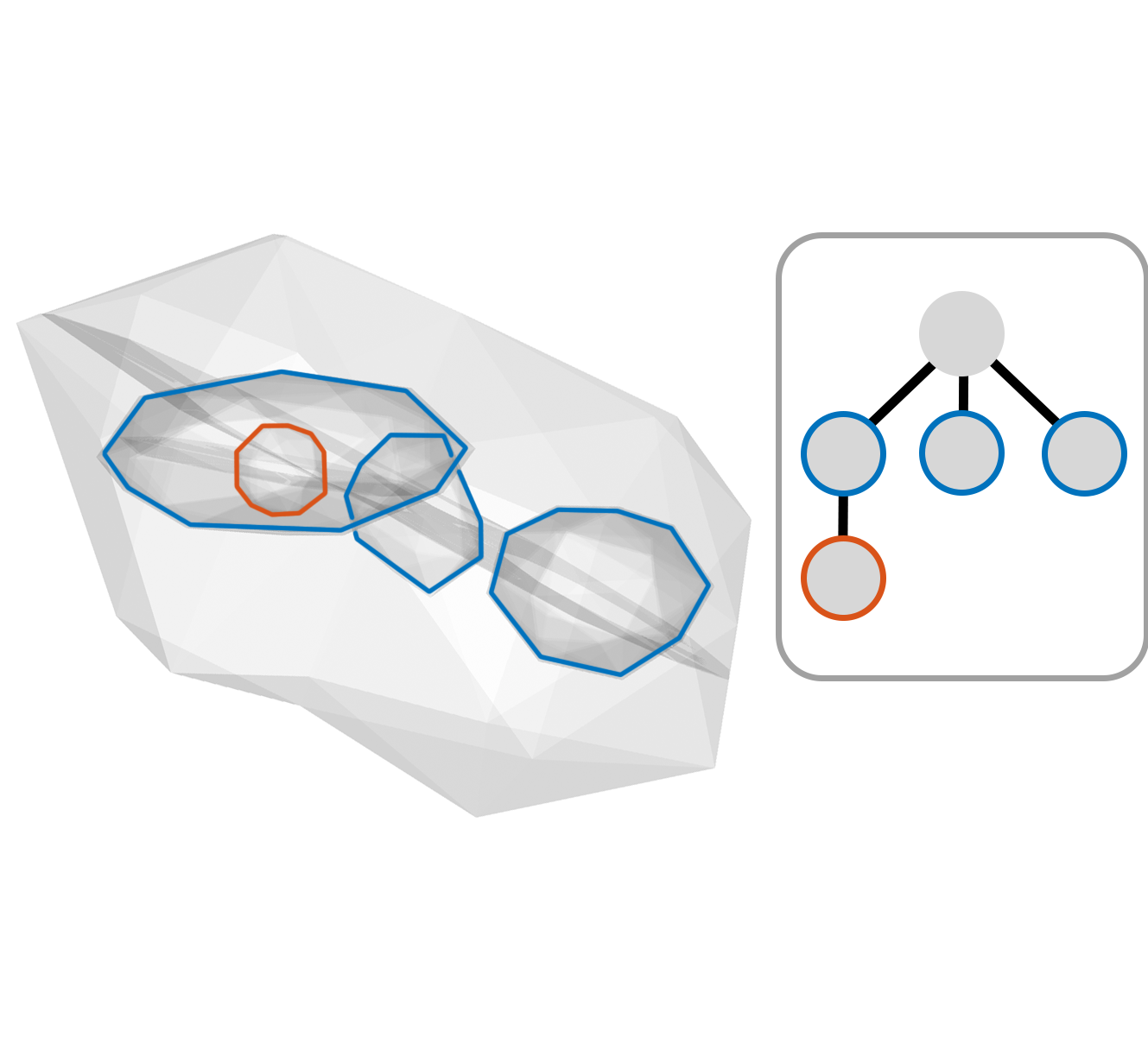}        
        \\
        (a) Synthetic Papermesh 1 & (b) Synthetic Papermesh 2
    \end{tabular}
    \caption{Two synthetic models used to experiment with different nested configurations.}
    \label{fig:result_synthetic_12}
    }
\end{figure}

\begin{figure*}[t]
    \centering{
    \setlength{\tabcolsep}{0pt}
    \begin{tabular}{cccc}
        \includegraphics[width=0.23\linewidth]{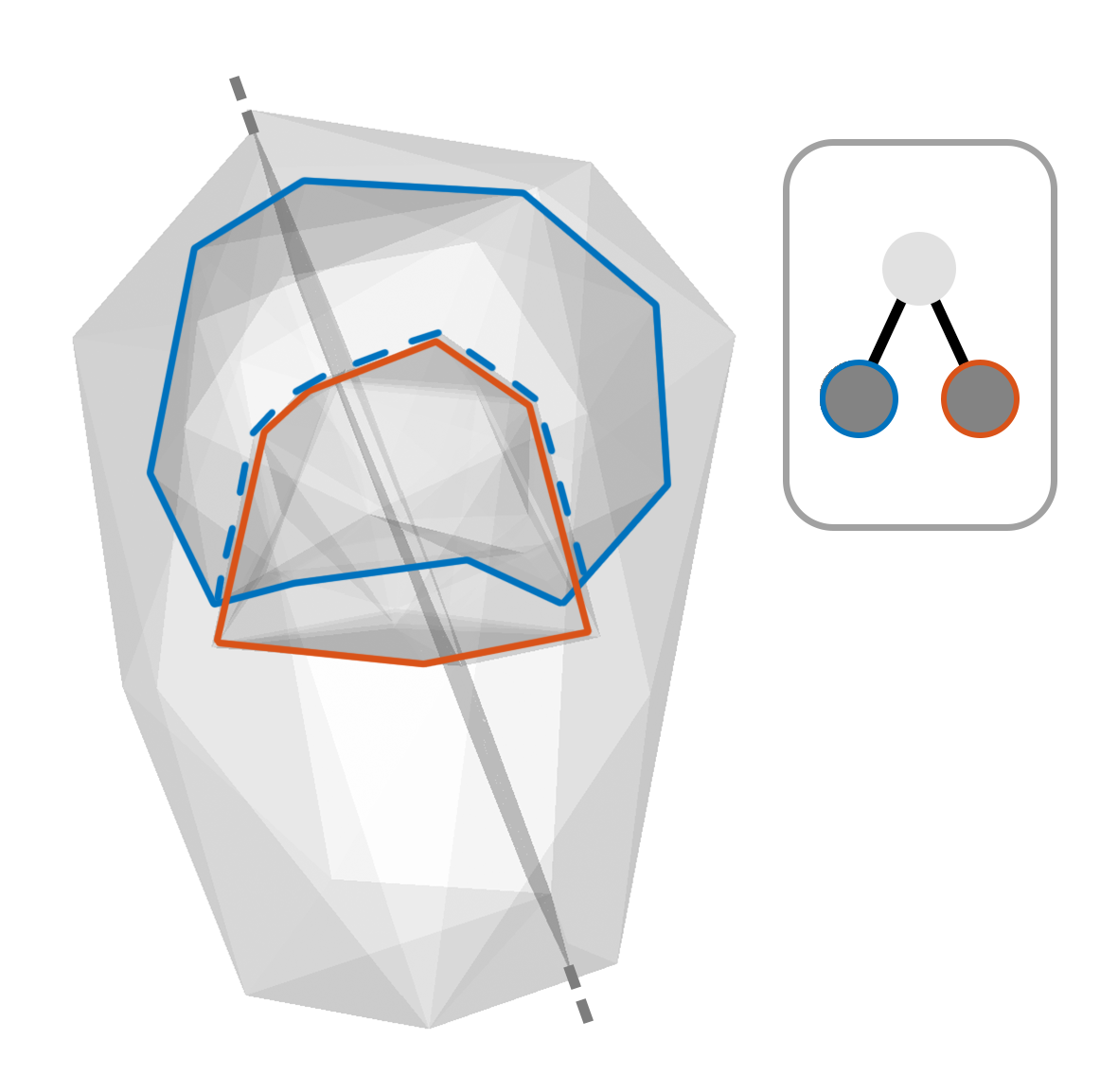} &
        \includegraphics[width=0.23\linewidth]{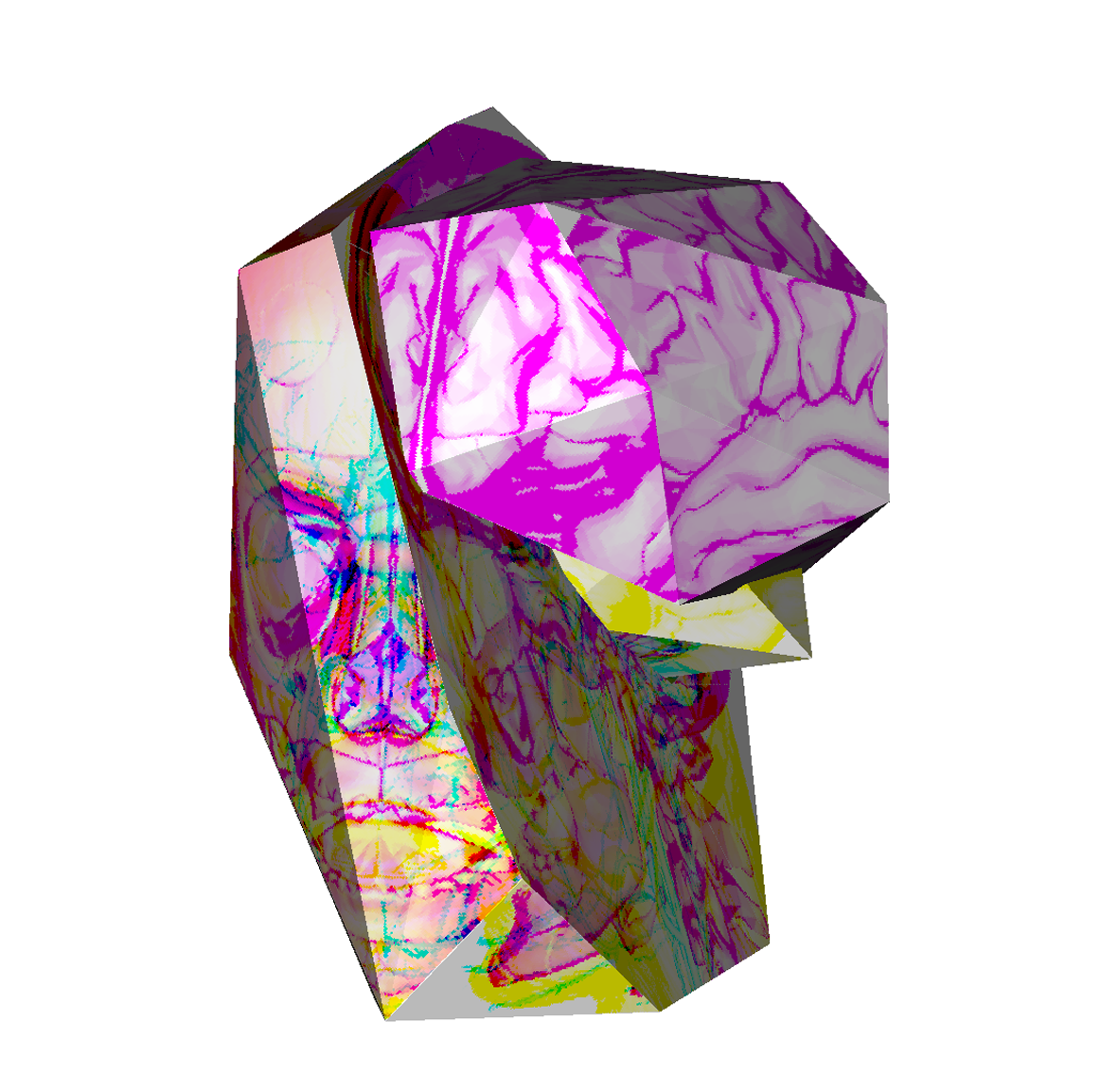} &
        \includegraphics[width=0.23\linewidth]{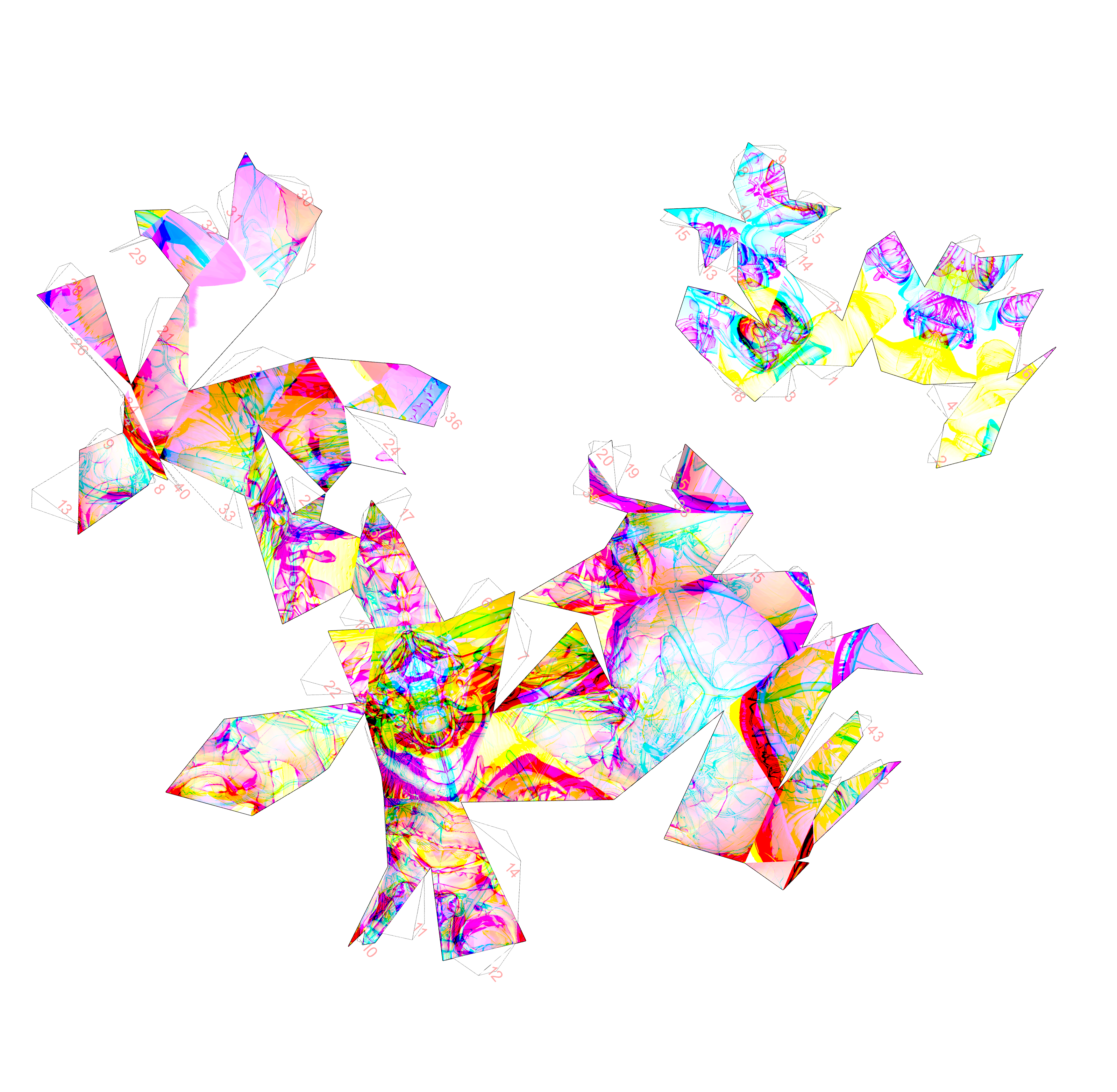} &
        \includegraphics[width=0.23\linewidth]{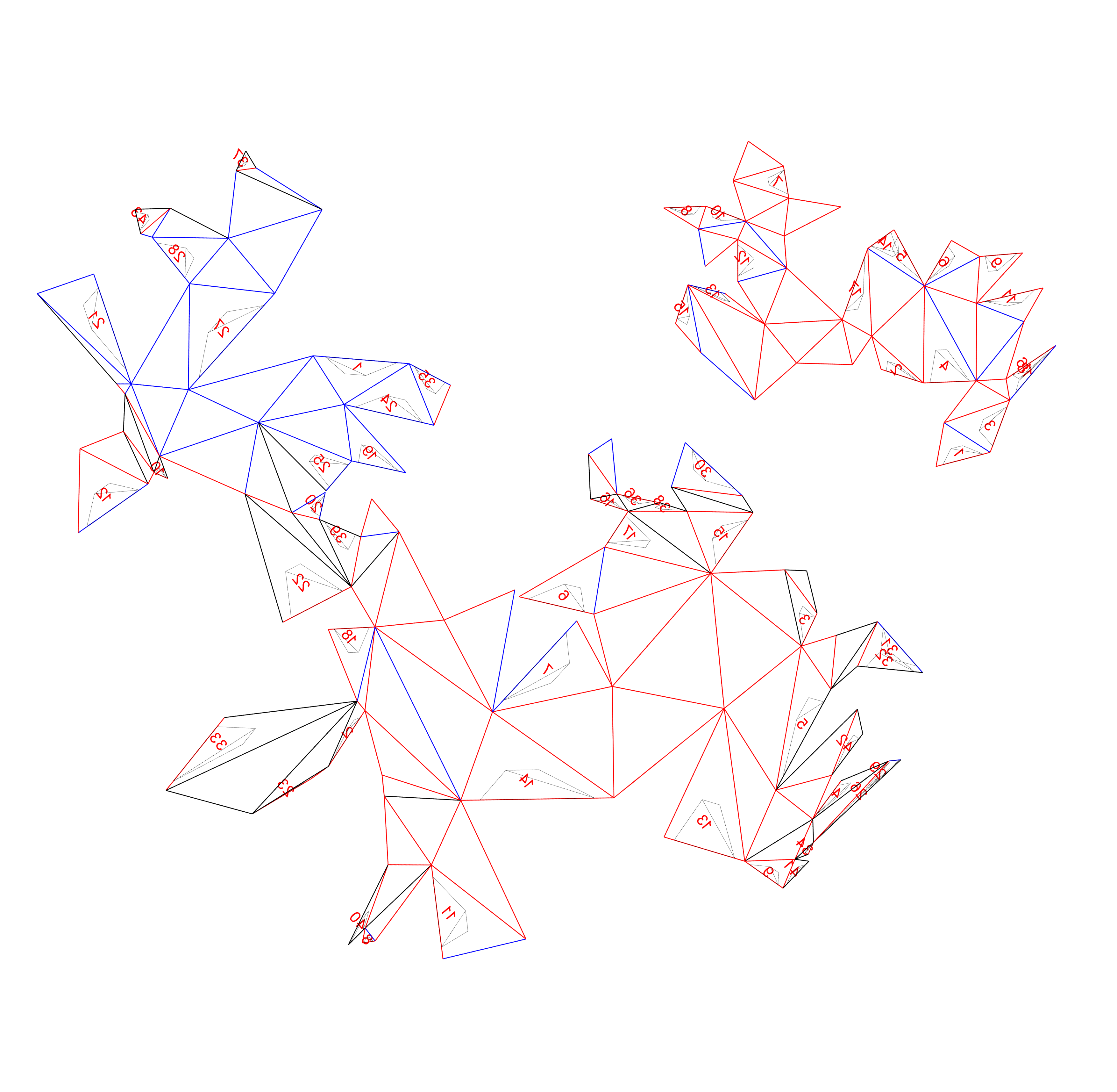} \\
        (a)  & 
        (b)   & 
        (c-1)  & 
        (c-2)  \\
       
        \includegraphics[width=0.23\linewidth]{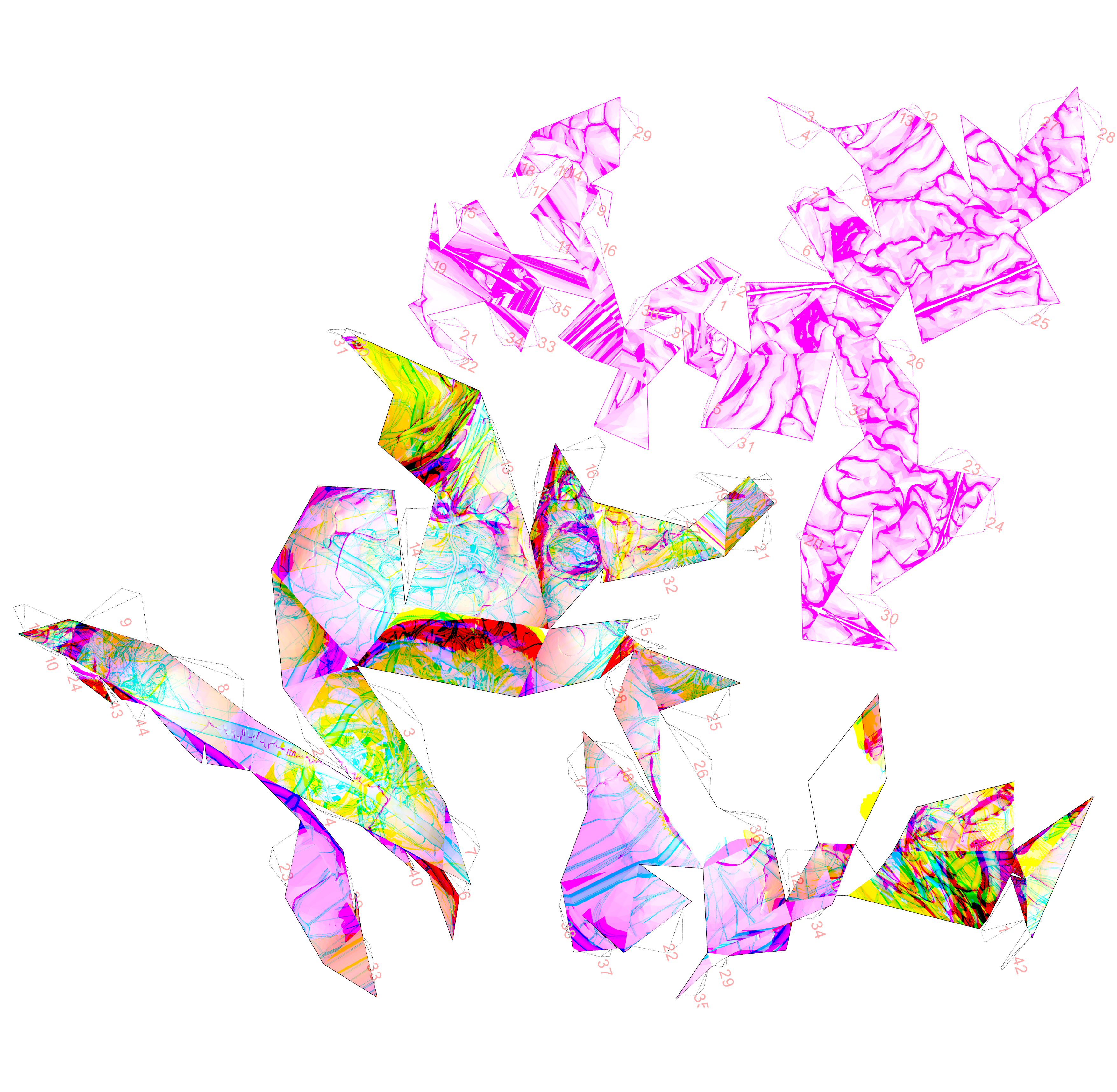} &        \includegraphics[width=0.23\linewidth]{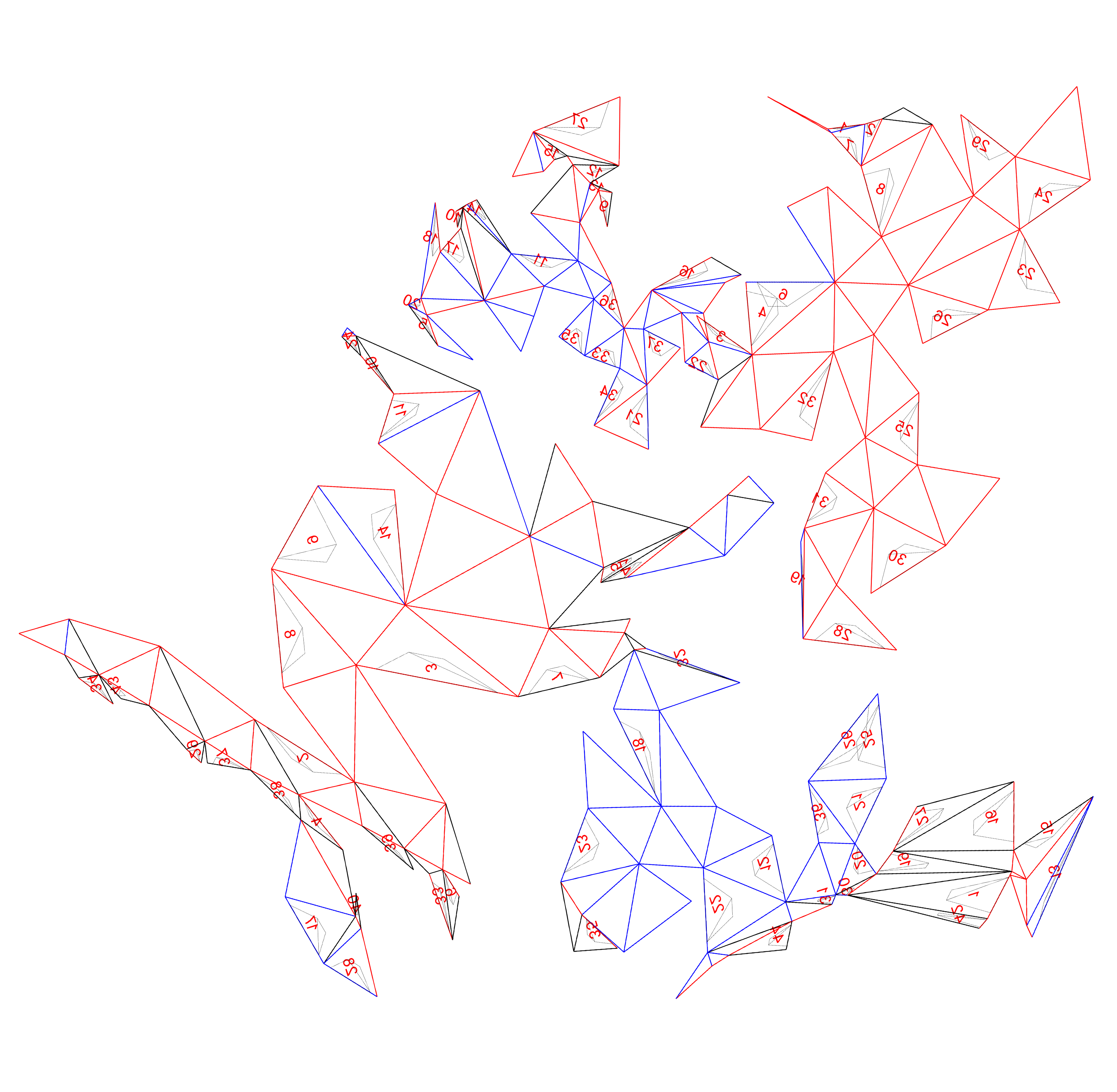} &
        \includegraphics[width=0.23\linewidth]{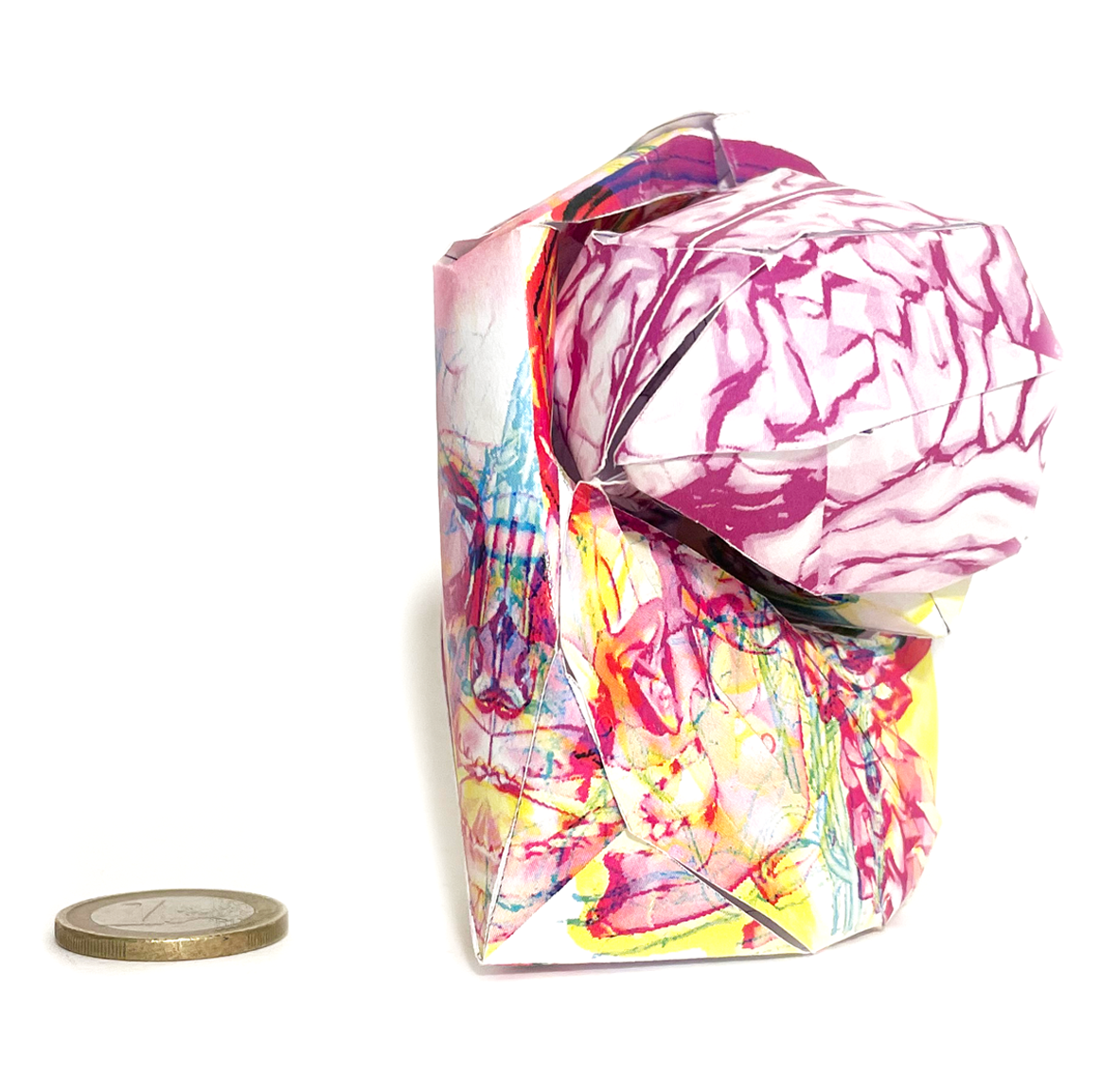} &        \includegraphics[width=0.23\linewidth]{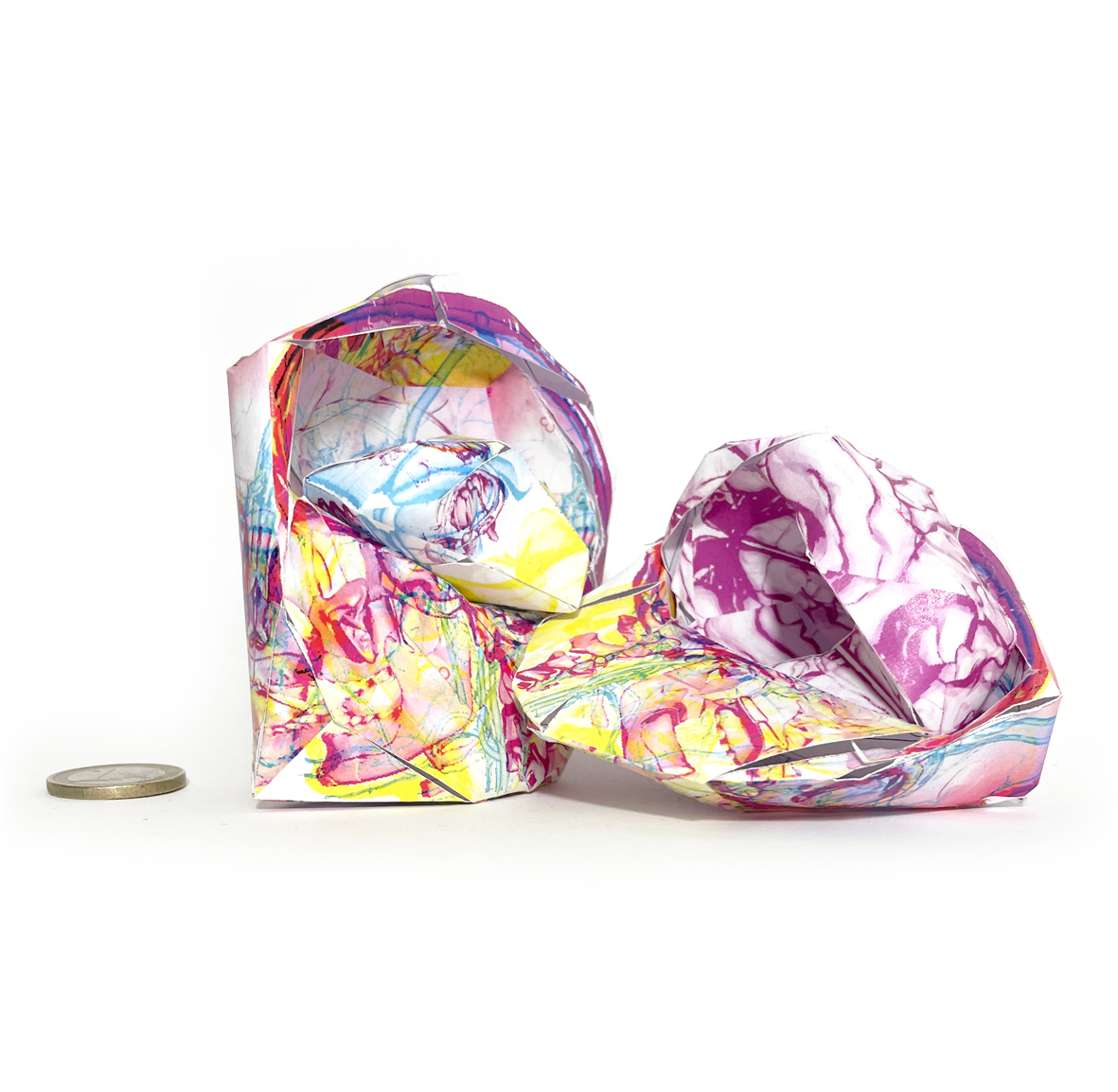} \\
        
        (c-3)   & 
        (c-4) & 
        (d) &
        (e) \\
    \end{tabular}
    \caption{Our workflow applied on a nested head model. The model was obtain from BodyParts3D/Anatomography~\cite{BP3D}. (a) The papermesh tree and cutting configuration. (b) Projected texture on papermeshes. (c-1)--(c-4) Unfolded patches for double-sided printing. (d) Assembled and (e) open papercraft, where the brain has been removed. \vspace{-8pt}}
    \label{fig:result_real_1}
    }
\end{figure*}
\vspace{-8pt}

\begin{figure*}[t]
    \centering{
    \setlength{\tabcolsep}{0pt}
    \begin{tabular}{cccc}
        \includegraphics[width=0.24\linewidth]{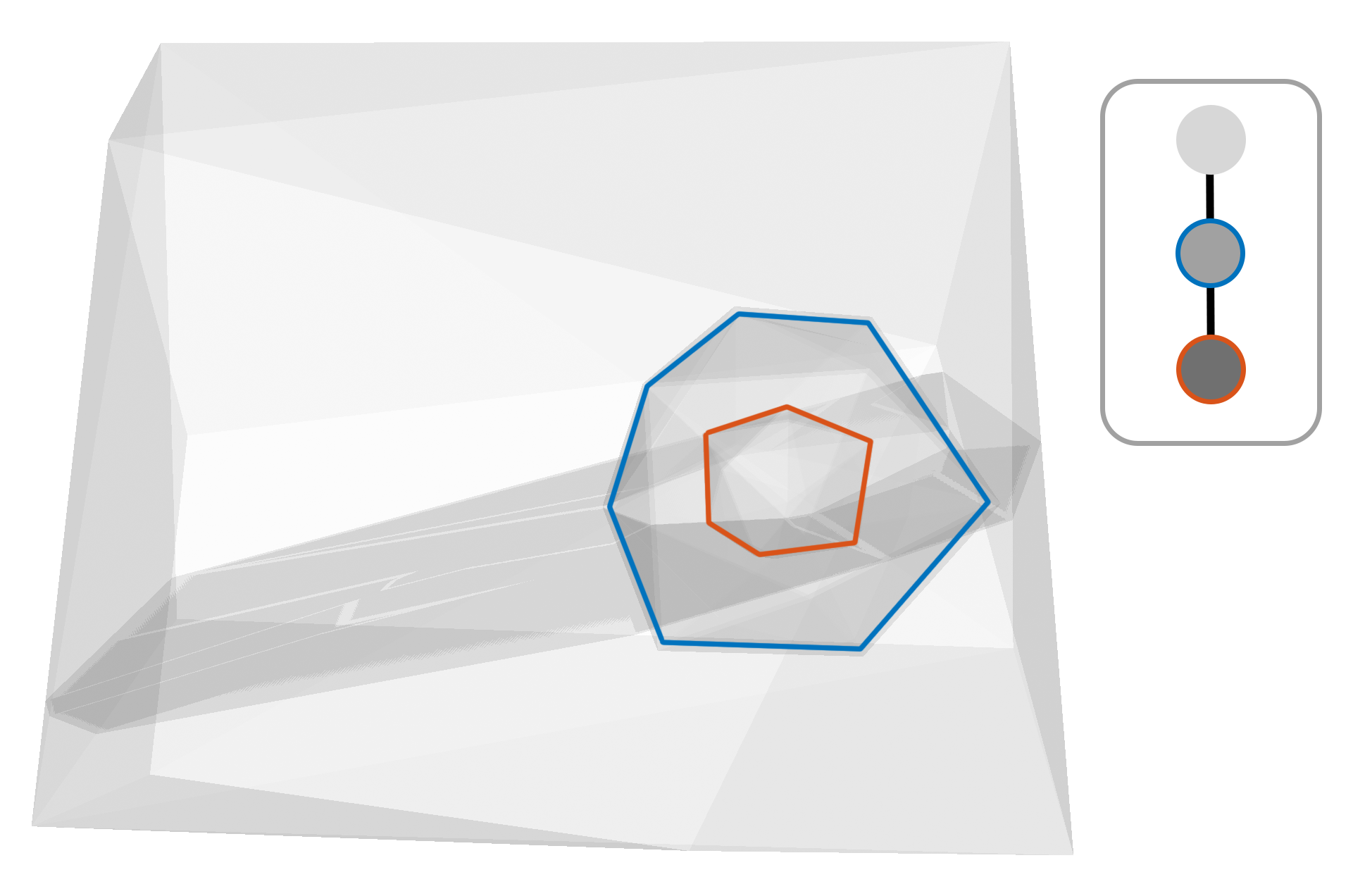} &
        \includegraphics[width=0.24\linewidth]{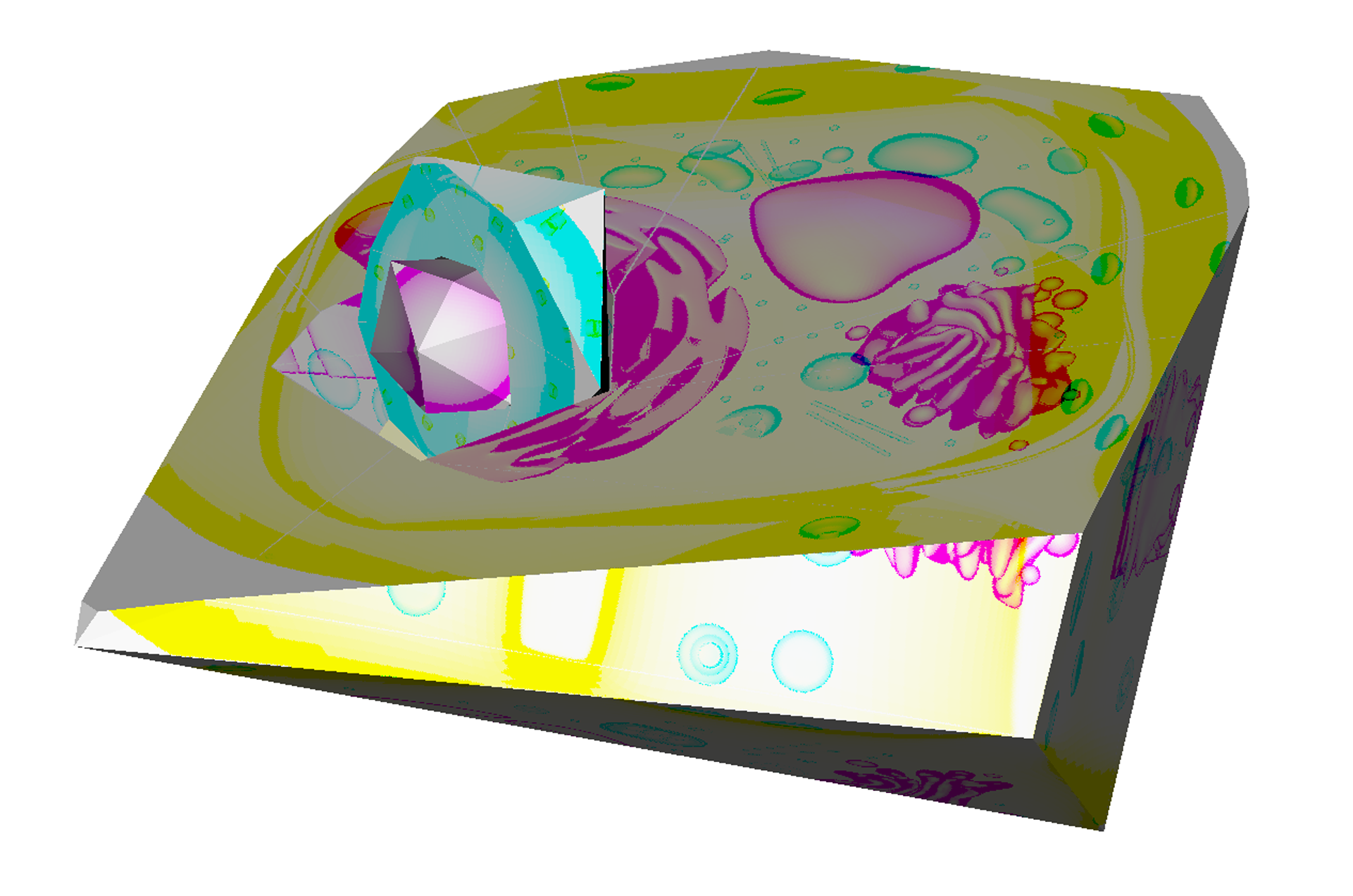} &
        \includegraphics[width=0.24\linewidth]{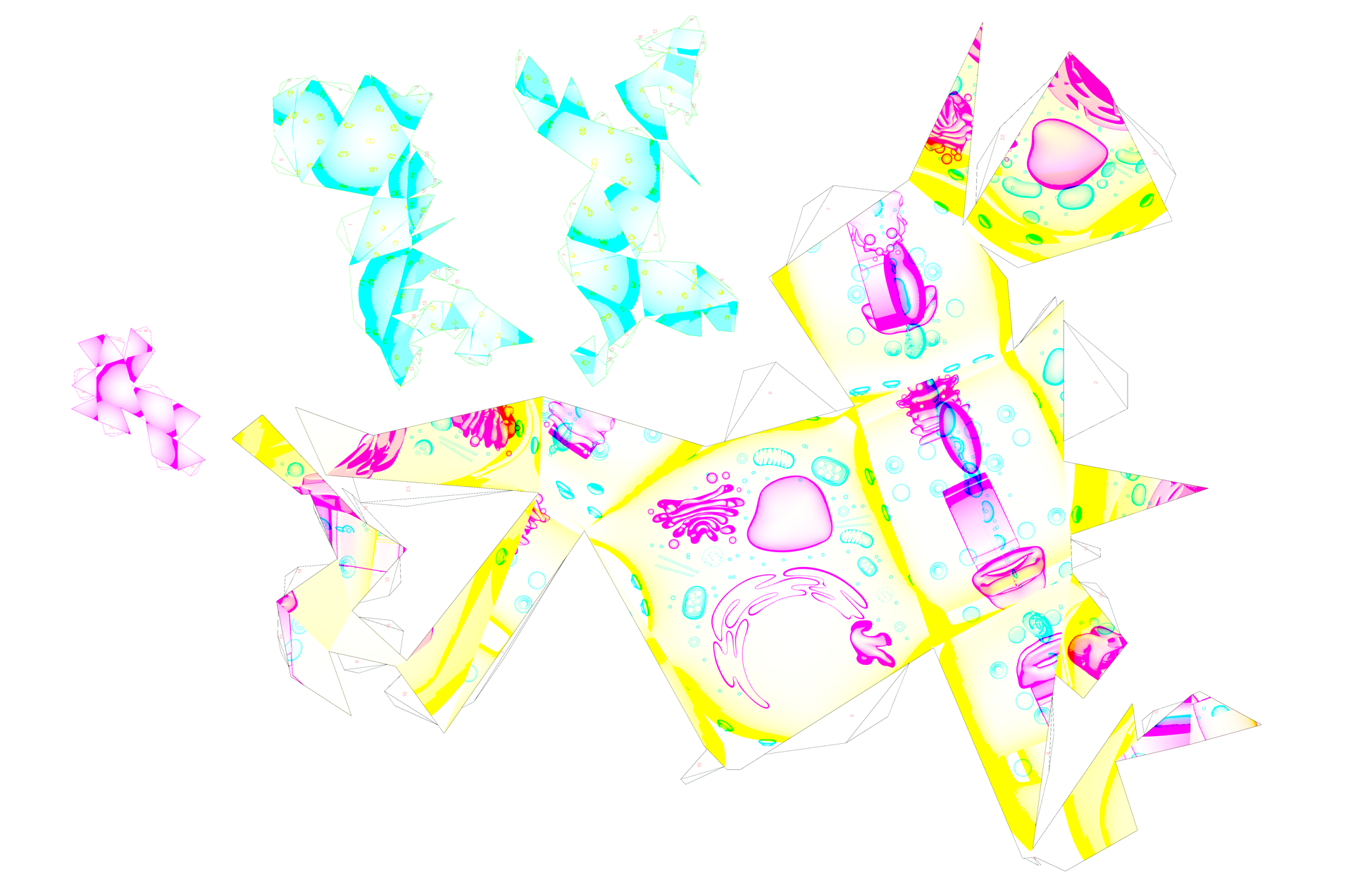} &
        \includegraphics[width=0.24\linewidth]{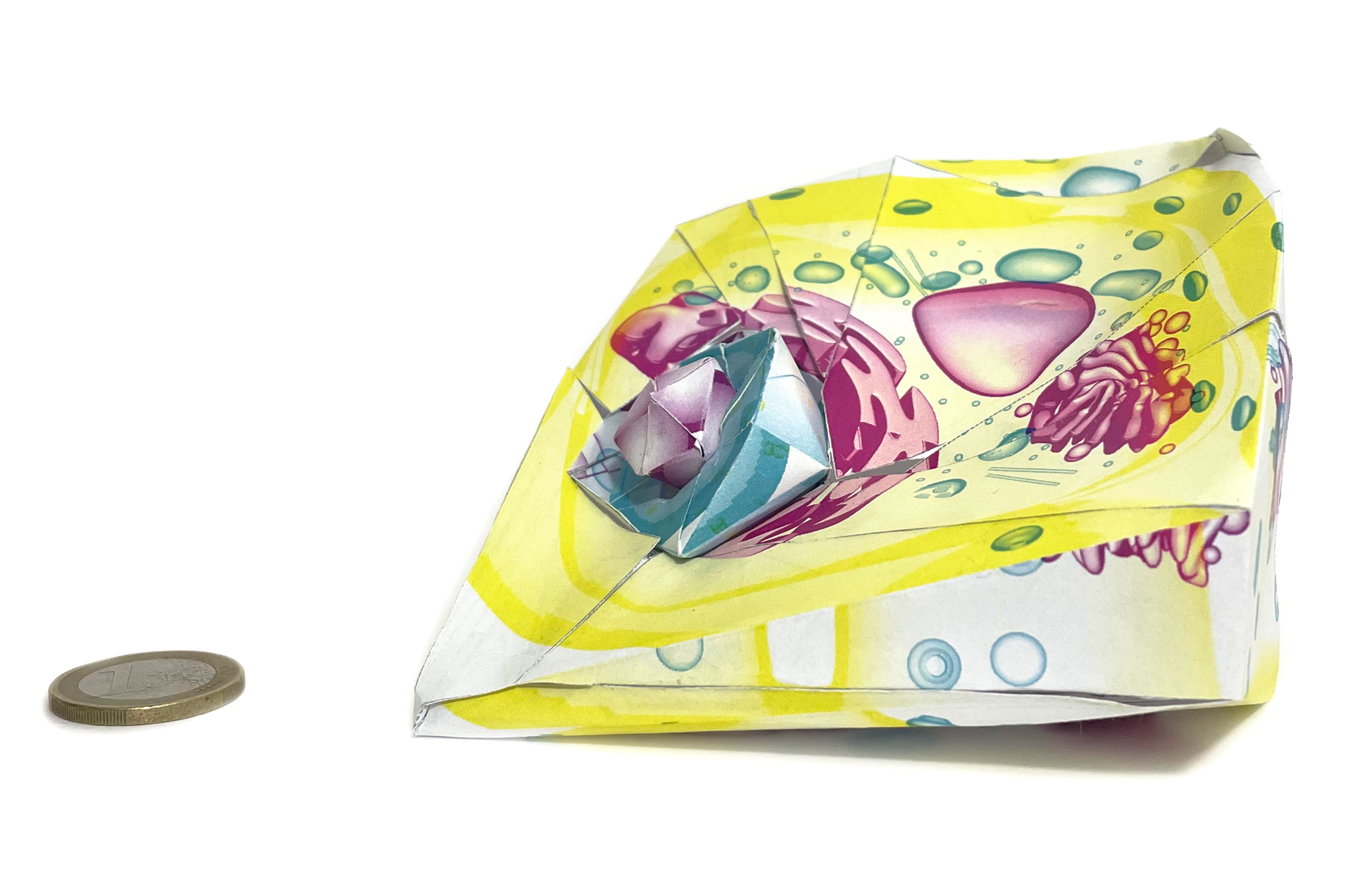} \\
        (a)   & 
        (b)  & 
        (c) & 
        (d) \\
         \\
    \end{tabular}
    \caption{Our workflow applied on a nested plant cell model. (a) The papermesh tree and cutting configuration. (b) Projected texture on papermeshes. (c) Unfolded patches for double-sided printing (only front side visible). (d) Assembled papercraft.}
    \label{fig:result_real_2}
    }
    \centering{
    \setlength{\tabcolsep}{0pt}
    \begin{tabular}{ccccccc}
        \includegraphics[width=0.14\linewidth]{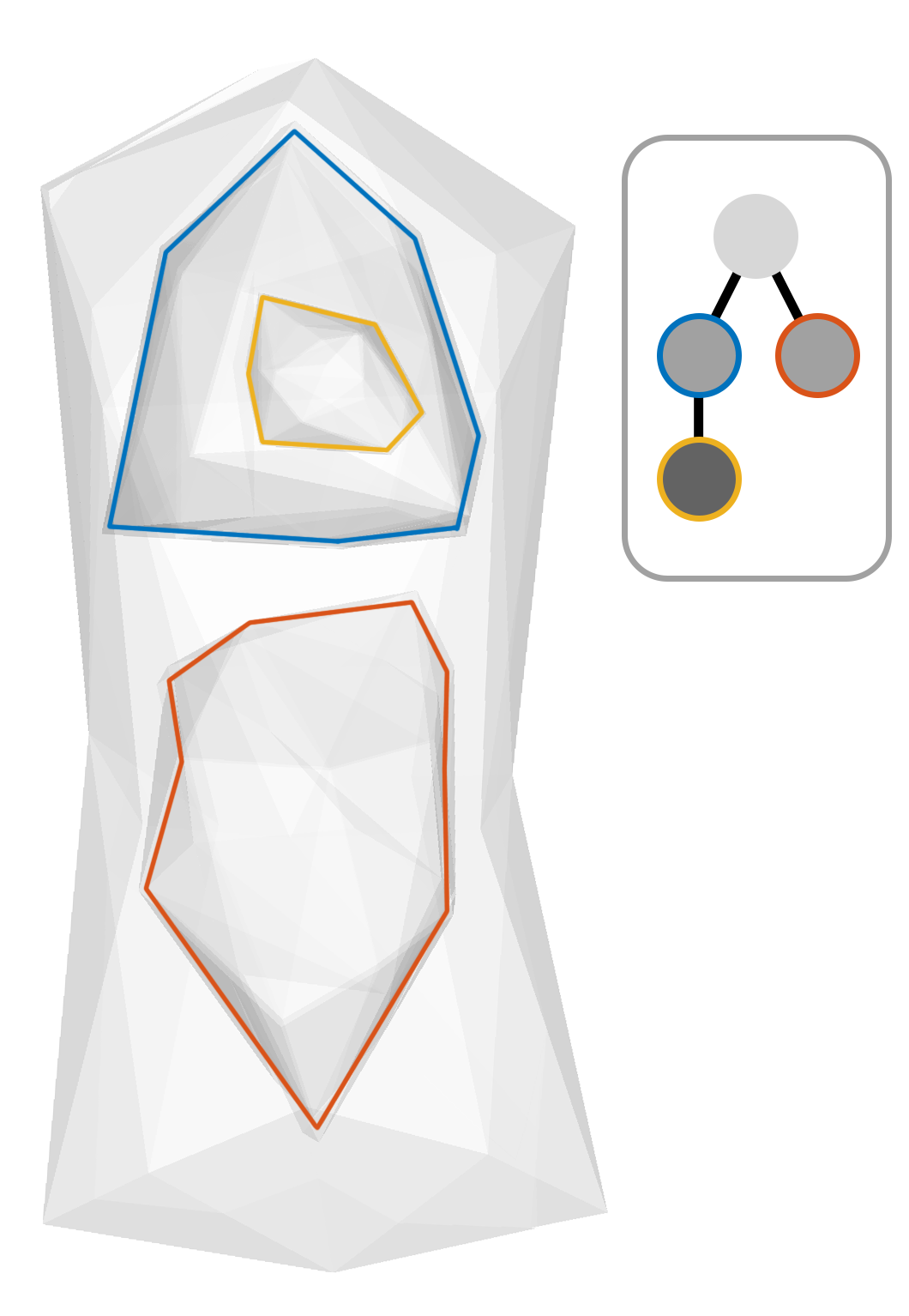} &
        \includegraphics[width=0.14\linewidth]{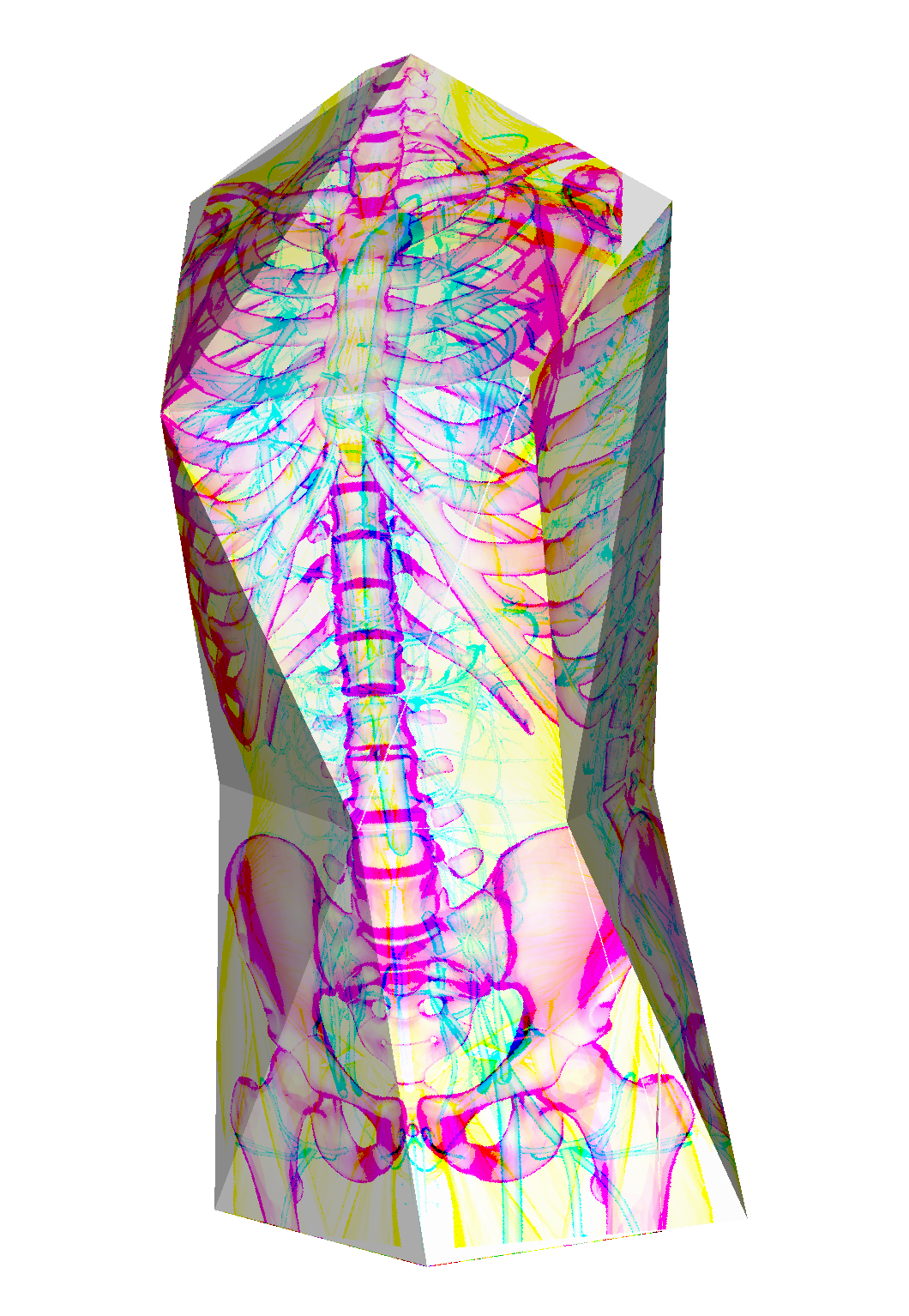} &
        \includegraphics[width=0.14\linewidth]{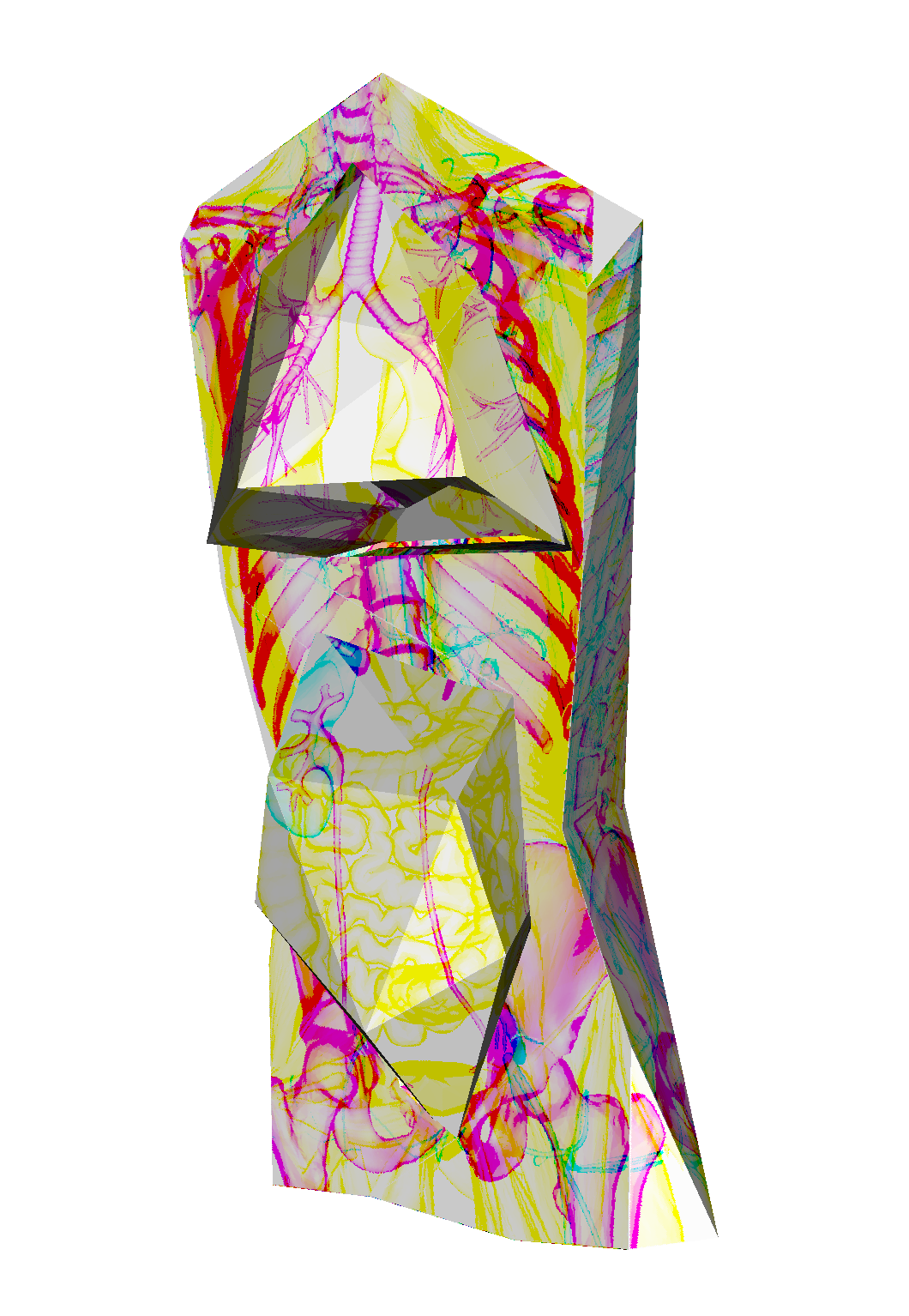} &
        \includegraphics[width=0.14\linewidth]{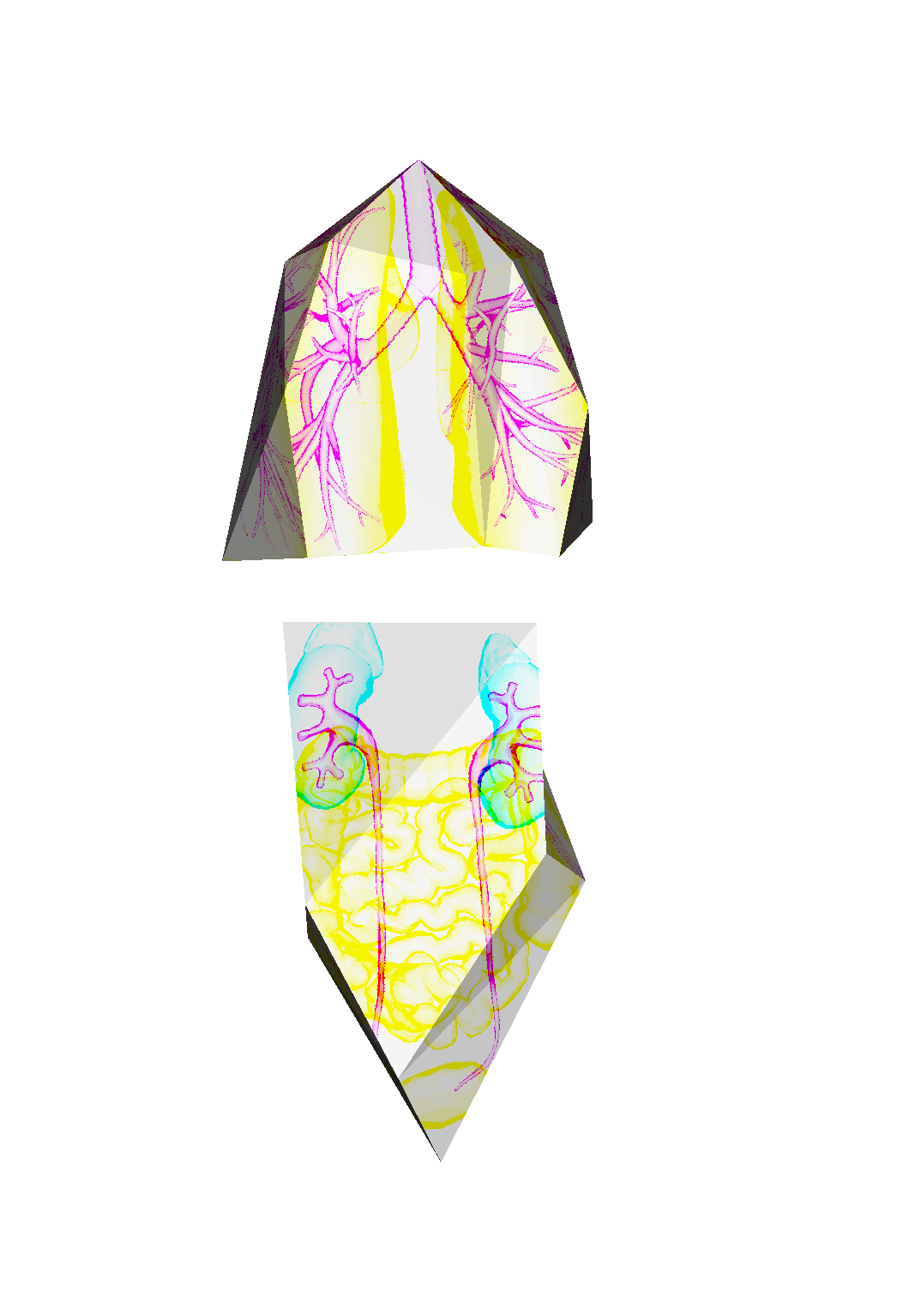} &
        \includegraphics[width=0.14\linewidth]{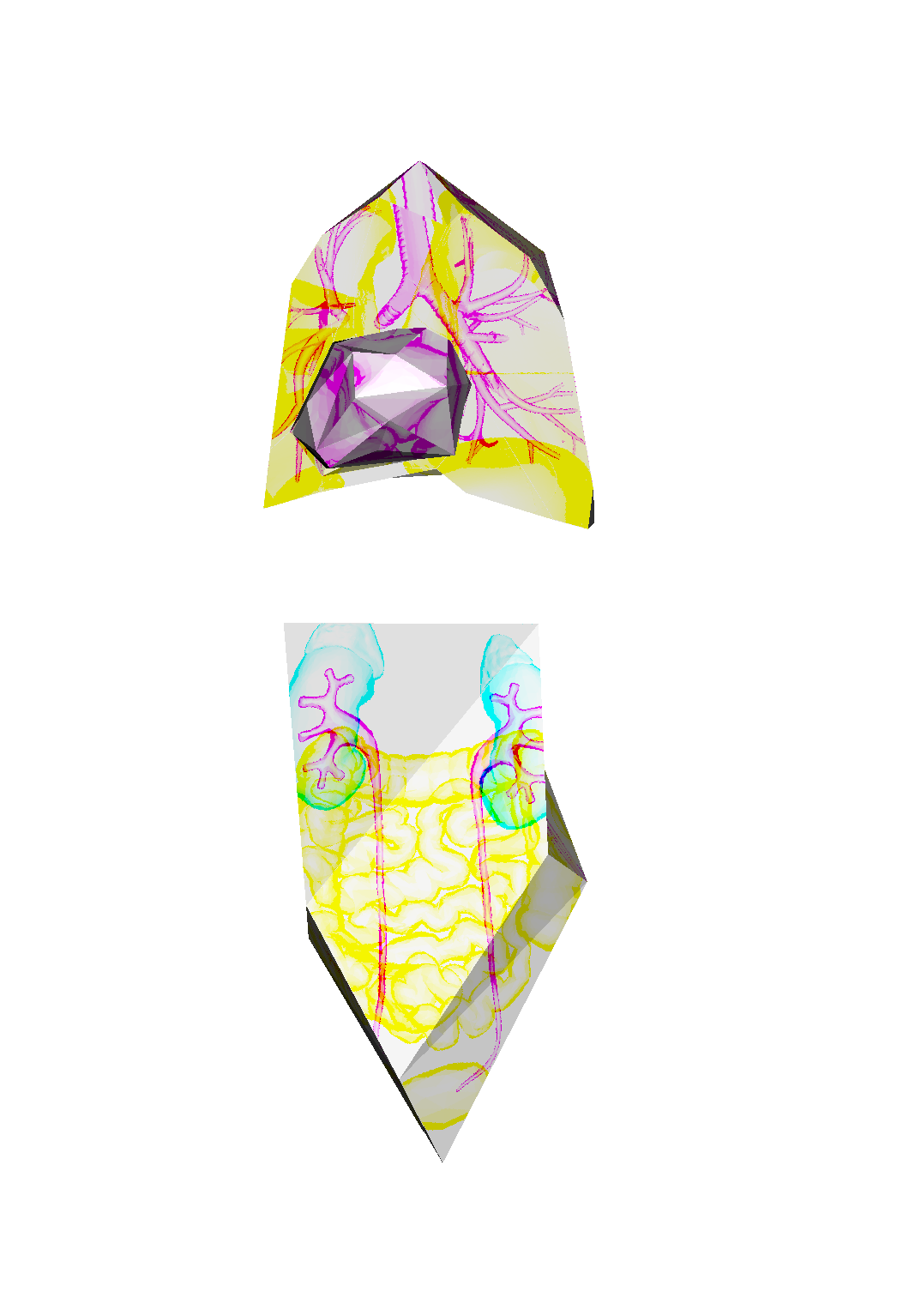} &
        \includegraphics[width=0.14\linewidth]{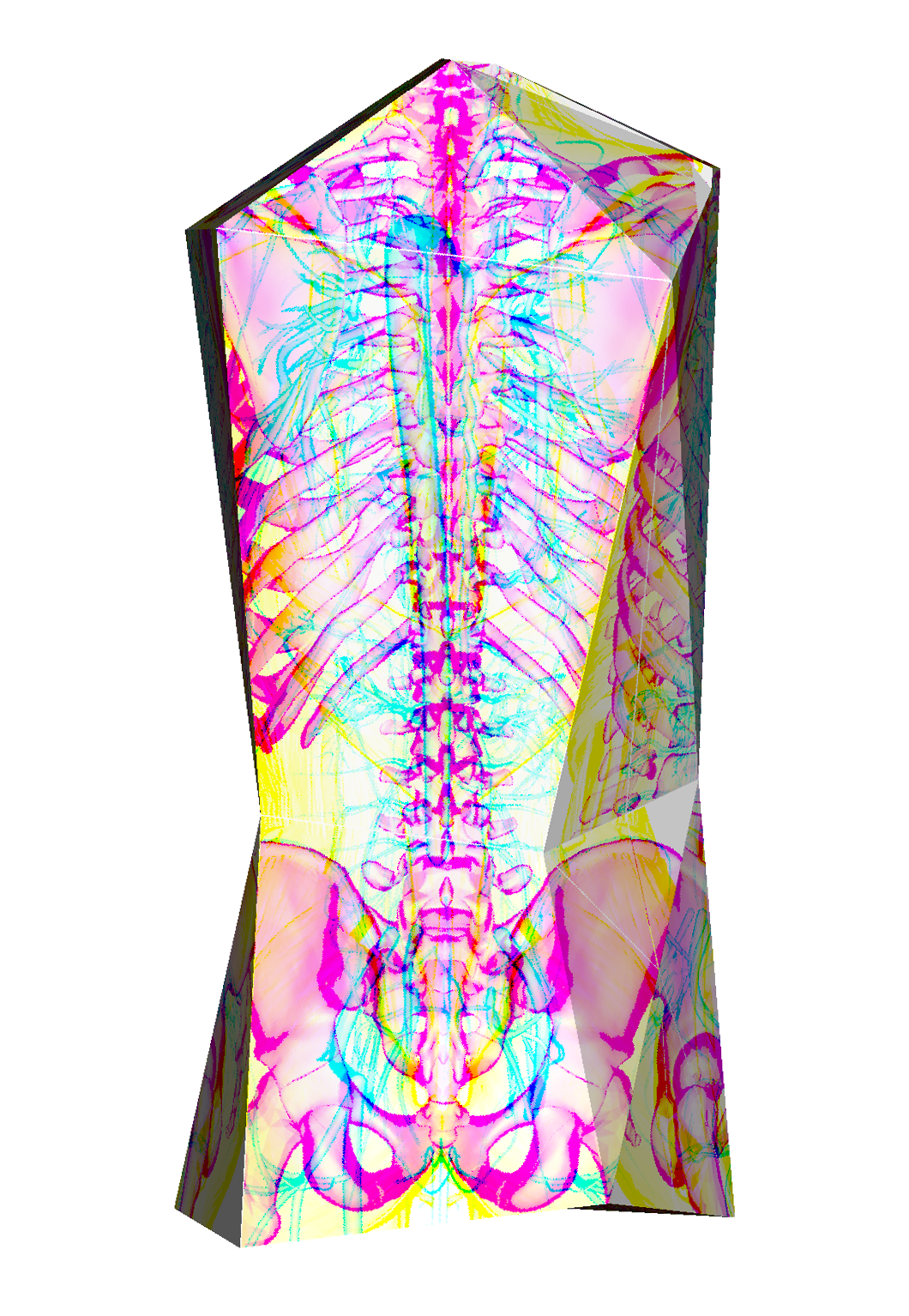} &
        \includegraphics[width=0.14\linewidth]{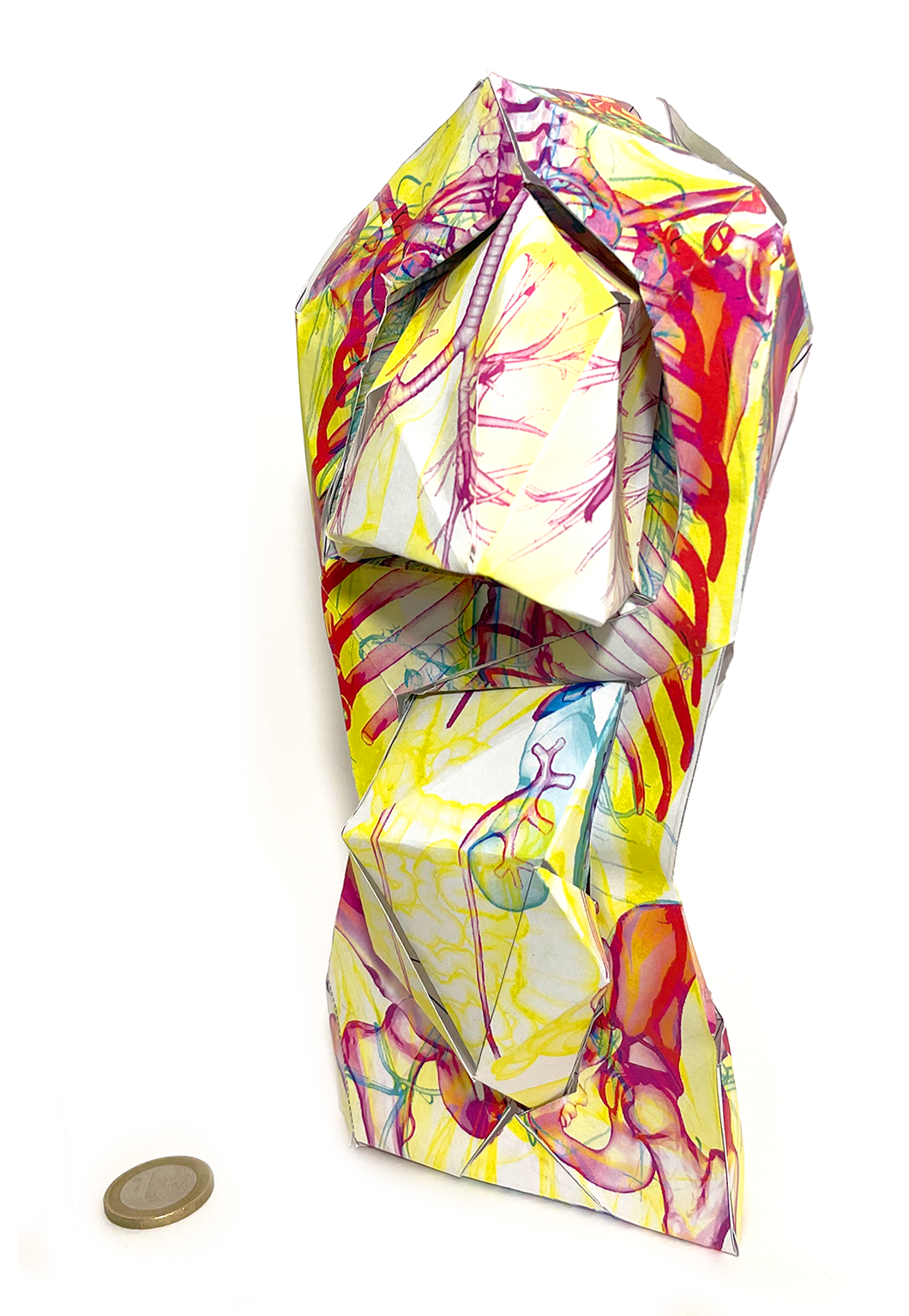}\\
        (a)   & 
        (b)  & 
        (c) & 
        (d)  & 
        (e)  & 
        (f) & 
        (g) 
        \\

    \end{tabular}
    \caption{Our workflow applied on a nested human torso model. The model was obtain from BodyParts3D/Anatomography~\cite{BP3D}. (a) The papermesh tree and cutting configuration. (b) Projected texture on the papermesh of the outer level. (c) Outer level open and middle level revealed. (d) Projected texture on the middle level. (e) Middle level open and inner level revealed (on the upper model, inner level corresponds to the heart). (f) Projected texture on the papermesh of the outer level (back side). (g) Assembled, open papercraft.\vspace{-5pt}}
    \label{fig:result_real_3}
    }
\end{figure*}

\section{Experimental Results}
\label{sec:result}

In this section, we demonstrate several nested papercraft examples generated with our approach. 
We first employ a set of synthetic models, where we experiment with nested structures of different topologies and hierarchies (Section~\ref{ssec:synthetic}). 
Then, we demonstrate several cases with anatomical and biological models (Section~\ref{ssec:real_data}). 
Table~\ref{table:meshstats} 
shows the number of vertices and faces for the input data and for the papermeshes (in parentheses), used in our examples.

\begin{table*}
\caption{
The total vertex and face count for input meshes and approximate papermeshes (in parentheses) of each 3D model.
}
\label{table:meshstats}
\centering{
\begin{tabular}{|c|cccccc|} 
\hline
Models & Head (Fig.~\ref{fig:teaser}) & Synthetic 1 (Fig.~\ref{fig:result_synthetic_12}(a)) & Synthetic 2 (Fig.~\ref{fig:result_synthetic_12}(b)) & Head (Fig.~\ref{fig:result_real_1}) & Plant cell (Fig.~\ref{fig:result_real_2}) & Torso (Fig.~\ref{fig:result_real_3}) \\
\hline
Vertices & 1,051,155 (102) & 252 (109) & 210 (152) & 1,051,155 (84) & 2,396 (60) & 1,722,665 (98)\\
Faces & 2,091,807 (196) & 480 (210) & 400 (296)  & 2,091,807 (156) & 2,316 (112) & 3,418,171 (188) \\
\hline
\end{tabular}
\vspace{-5pt}}
\end{table*}
\vspace{-5pt}

\subsection{Synthetic Models}
\label{ssec:synthetic}
Synthetic meshes are used to showcase the theoretical possibilities of the pipeline, disregarding the possibility of reconstruction due to the limited paper size, when printing unfoldings.
Figure~\ref{fig:result_synthetic_12} 
shows two different complex configurations of nested meshes. 
In these cases, we focus on the hierarchy detection, the viewpoint calculation, the identification of the optimal cutting plane, and the stability calculation (i.e., steps (a)--(f) in Figure~\ref{fig:overview}).
Our two examples demonstrate that the first part of the workflow is possible and gives solutions within reasonable times.
The reconstruction of such complex models might not be possible, due to the limited size of printed patches which makes the nested structures too small to assemble.
\vspace{-5pt}

\subsection{Anatomical and Biological Nested Models}
\label{ssec:real_data}
Figure~\ref{fig:teaser} shows the workflow steps to obtain results of a nested papercraft for a \textit{human head}, with only two nested structures, i.e., the outer head and the inner brain. 
Figure~\ref{fig:result_real_1} shows a human head papercraft with a more complex nested configuration. 
Here, the inner level is split into two subparts, as shown in the tree and cutting configuration (Figure~\ref{fig:result_real_1} (a)).
The outer model consists of muscles (in magenta), bones (in cyan), and veins (in yellow), projected onto the papermesh (Figure~\ref{fig:result_real_1} (b)), the unfolding (Figure~\ref{fig:result_real_1} (c-1,3)) and papercraft (Figure~\ref{fig:result_real_1} (d,e)) surface. 
The inner model consists of two meshes approximating the forebrain (in magenta), and the midbrain together with the hindbrain as nested structures of the head.
These are also projected on the papermesh (Figure~\ref{fig:result_real_1} (b)), the unfolding (Figure~\ref{fig:result_real_1} (c-1,3)) and papercraft (Figure~\ref{fig:result_real_1} (d,e)) surface. 
This shows the typical use case for projecting structures that cannot be turned into papermeshes, while enabling the exploration of bigger structures and their relation to each other as papermeshes.

Figure~\ref{fig:result_real_2} shows a nested papercraft example for a \textit{plant cell}. 
The model is split into three levels in the tree configuration (Figure~\ref{fig:result_real_2} (a)).
Various small or complex structures, i.e., plasmodesmata, amyloplast, peroxisome and the Golgi apparatus (either in magenta or in cyan) are projected on the outer level that represents in yellow the cytoplasm (Figure~\ref{fig:result_real_2} (b,c)). 
The nucleus and nucleolus can be unfolded as individual nested papermeshes (see respectively cyan and magenta in Figure~\ref{fig:result_real_2} (b,c)), and assembled (Figure~\ref{fig:result_real_2} (d)). 
\rv{Assembling the outer level, visible in Figure \ref{fig:result_real_2} (c) took 28 min 49 s, excluding the time for cutting out the template.}

Figure~\ref{fig:result_real_3} shows a nested papercraft of the \textit{human upper body}. 
The tree configuration results into three levels for this model (Figure~\ref{fig:result_real_3} (a)).
The outer level shows the veins (in cyan), muscles (in magenta) and bones (in yellow) projected onto the surface (Figure~\ref{fig:result_real_3} (b,c,f)). \rv{The papercraft for the outer level, visible in Figure \ref{fig:result_real_3} (g) took 41 min 58 s to assemble.}
The middle level contains the intestines (in yellow), bladder (in magenta) and kidneys (in cyan) together, and separately the lungs (in yellow) with the bronchial tree (in magenta) (Figure~\ref{fig:result_real_3} (d,e)). 
The last level contains the heart (Figure~\ref{fig:result_real_3} (e)). In this case the nested meshes also have multiple projected structures.
Projecting all of the present structures onto the same surface would lead to visual clutter, making it harder to understand the distinct relation between different structures. 
Instead, generating multiple papermeshes for important organs leads to less visual clutter without losing information (Figure~\ref{fig:result_real_3} (g)).
All of the real world examples were printed on A3, and the relative scale is indicated on the figures for each of the models. 
\vspace{-5pt}
\section{Evaluation and Discussion}
\label{ssec:discuss}

In this section, we assess the feasibility of the presented approach. 
This includes a performance test of the components of the workflow (Section~\ref{ssec:performance}) and a feasibility test for papercraft reconstruction (Section~\ref{ssec:feasibility}). 
This is followed by an informal interview with users
(Section~\ref{ssec:interview}).
\rv{We conducted also a second study to evaluate the impact of mountain-valley folds on the papercraft assembly (Section~\ref{ssec:mountain-valley-folds}).}
We include also a discussion of the current limitations and intuitive solutions thereof (Section~\ref{ssec:limit}). 

\vspace{-5pt}
\subsection{Quantitative Evaluation}
\label{ssec:performance}

Our system was tested on a Windows 10 machine with a 3.8 GHz, 64MB L3 Cache Processor and 32GB of RAM. 
Table~\ref{table:runtime} gives an overview of the running time in seconds of each component of our workflow.
The first steps (components (a)--(f) of Figure~\ref{fig:overview}, as discussed in Sections~\ref{ssec:input}--~\ref{ssec:stability}) do not take too much time. 
The primary computational bottlenecks come from the projection (Section~\ref{ssec:projection}, Figure~\ref{fig:overview}(g)) and unfolding (Section~\ref{ssec:unfolding}, Figure~\ref{fig:overview}(h)). 
The former is an iterative approach~\cite{schindler2020anatomical} and the latter is an optimization approach~\cite{Korpitsch:2020:WSCG}, which justifies their performance times.

\begin{table*}
\vspace*{0mm}
\small
\caption{
\vspace{-5pt}The running time (in $s$) of the results showcased in this paper. \rv{The last column indicates the total time (in $s$ and in $mm:ss$).}}
\label{table:runtime}
\centering{
\begin{tabular}{|p{22pt}|ccccccc|c|} 
\hline
&
\includegraphics[width=0.08\linewidth]{figures/workflow-2-hull.pdf} &
\includegraphics[width=0.08\linewidth]{figures/workflow-3-tree.pdf} &
\includegraphics[width=0.08\linewidth]{figures/workflow-4-viewpoint.pdf} &
\includegraphics[width=0.08\linewidth]{figures/workflow-5-cut.pdf} &
\includegraphics[width=0.08\linewidth]{figures/workflow-6-stability.pdf} &
\includegraphics[width=0.08\linewidth]{figures/workflow-7-projection.pdf} &
\includegraphics[width=0.08\linewidth]{figures/workflow-8-unfolding.pdf} &
\rv{\textbf{Total time}}\\
Step & Approximation & Hierarchy & Best viewpoing & Cutting & Stability & Projection & Unfolding & \rv{in $s$ (and in $mm:ss$)}\\
\hline \hline 
Fig.~\ref{fig:teaser}       & 2.04 & 0.05 & 0.95 & 0.3 & 0.484 & 461.1 & 6.69 & \rv{ 471.614 (7:52)} \\
Fig.~\ref{fig:result_real_1} & 2.23 & 0.1 & 1.07 & 0.54 & 0.486 & 482.41 & 40.11 & \rv{ 526.946 (8:47)} \\
Fig.~\ref{fig:result_real_2} & 9.85 & 0.05 & 1.22 & 0.56 & 0.502 & 727.25 & 19.78 & \rv{ 759.212 (12:39)} \\
Fig.~\ref{fig:result_real_3} & 2.87 & 0.12 & 1.17 & 1.23 & 0.501 & 863.28 & 627.01 & \rv{ 1496.181 (24:56)} \\
\hline
\end{tabular}
\vspace{-5pt}}
\end{table*}

\vspace{-5pt}
\subsection{Feasibility of Mesh Construction}
\label{ssec:feasibility}

We conducted a user study to investigate the feasibility of constructing the nested papercrafts. 
We recruited $10$ participants ($3$ females and $7$ males) with ages ranging from $25$ to $39$. 
\rv{The participants were mostly colleagues and students.} 
Only one participant (User 4, female) has no computer graphics background and one participant (User 7, male) is colorblind. 
We prepared the head model shown in Figure~\ref{fig:teaser}, being a model of adequate complexity and an interesting structure to learn anatomy.
This model includes three components: the left and right half of the head, and the brain inside the head. 
It is adequately challenging to construct, due to the concavity of half head papercraft, which contains several sharp angles in the papermesh.
The patches were printed double-sided on multiple \rv{A3 (220 GSM)} papers.
For each participant, we started by explaining the process of building individual papercrafts and pointed out that a single connected component corresponds to an enclosed papercraft. 
We moved on to explaining the meaning of the numbers on the glue tabs and that these glue tabs should be attached to the corresponding number on the back side using tape.
\rv{We instructed the participants to start with the small papercraft of the brain (28 patches), as this is easier and could serve also as an introduction to the assembly of the other, more complex structures.}
The participants were allowed to observe the patches for as long as they needed and to ask questions regarding the renderings before reconstruction.
This part was not included in the timings.
We note that once the time measurement began, the participants could use their own strategy to explore and construct the papercraft, as they would do in a real-life scenario.
Table~\ref{table:constructtime} summarizes the total construction time \rv{(excluding cutting time)} of the nested papercraft for each participant.
All participants managed to build the papercraft within less than two hours (average time was 90 min 41 s and standard deviation 21 min 22 s). 

\begin{table*}
\vspace*{0mm}
\small
\caption{
The assembly time (in $mm$:$ss$) of the head model (Figure~\ref{fig:teaser}) for each participant (F: female and M: male) \rv{for the two studies (A: without and B: with mountain-valley folds), excl. cutting time. We denote average (avg), standard deviation (sd), and no participation (N/A).}}
\label{table:constructtime}
\centering{
\begin{tabular}{|l|cccccccccc|c|} 
\hline
User & 1 (M) & 2 (M) & 3 (F) & 4 (F) & 5 (F) & 6 (M) & 7 (M) & 8 (M) & 9 (M) & 10 (M) & avg $\pm$ sd\\
\hline
Time A (mm:ss) & 99:38 & 95:04 & 112:32 & 99:12 & 59:04 & 53:02 & 98:04 & 103:07 & 117:04 & 67:32 & 90:26 $\pm$ 22:23 \\
\hline
\rv{Time B (mm:ss)} & \rv{85:52} & \rv{92:21} & \rv{99:56} & \rv{89:33} & \rv{48:52} & \rv{40:06} & \rv{86:23} & \rv{N/A} & \rv{N/A} & \rv{N/A} & \rv{77:35 $\pm$ 23:13 }\\
\hline
\end{tabular}
\vspace{-5pt}}
\end{table*}

\subsection{Interview with Evaluation Participants}
\label{ssec:interview}

After completing the reconstruction of the papercrafts, we asked the evaluation participants about their experience with the physicalization. 
When asked about whether the papercraft was easy to build or not, most of the participants indicated that they had difficulties---especially, when building the ``negative'' side of the head level, i.e., the head--brain interface. 
Reconstructing just the brain was considered easy by most, and all reconstructed it fast. 
Most people would have appreciated additional indicators for the reconstruction. 
One person commented that they ``\textit{would like to build one with a single texture, to see if it would be easier}''.
We also asked the participants what they consider to be an upper limit in number of meshes for a feasible reconstruction. 
\rv{Most participants commented that 3--5 meshes would be the upper limit for them.}
We also asked them about a comparison to traditional on-screen visualizations. 
One participant commented that ``\textit{it is interesting to be able to take the brain out with [their] own hands}'', and another one mentioned that ``\textit{[..] the scope [i.e., between physicalization and visualization] is different}''.
Two people were clearly in favor of an on-screen visualization.
When asked about particular findings, 
one person commented that ``\textit{you can see a very surprising scale of how big the brain is in comparison to the skull and that is more noticeable than a 3D rendering}''.
We finally asked them about potential applications, and half of the participants commented about children education. 
One commented that the model is ``\textit{very fragile to use for anything else than entertainment [and] would seem unprofessional to use in patient communication}''.
 
\vspace{-5pt}
\subsection{\rv{Effect of Mountain-Valley Folds on Assembly}}
\label{ssec:mountain-valley-folds}

\rv{In a second study, we assess the effect of mountain-valley folds on the papercraft assembly. 
Two months after the first study, we asked the same people to participate in a second session. 
Participants 8--10 were not available, but participants 1--7 underwent the same process as described in Section~\ref{ssec:feasibility}. 
The difference in this case was that the papercrafts had an additional encoding of the mountain-valley fold~\cite{takahashi2011optimized}, supporting the users in folding the patch towards the ``outside'' or the ``inside''}.
\rv{Again, we measured the assembly time for all participants.
The results are summarized in the last row of Table~\ref{table:constructtime}, and they confirm that the assembly time decreases with the use of mountain-valley folds by more than about 13 minutes (average time was 77 min 35 s and standard deviation 23 min 13 s).
Other smart guidance techniques to support the efficient assembly of the papercrafts should be further investigated, while coupling with informative aids could make the process more engaging.}
\vspace{-5pt}

\subsection{Limitations}
\label{ssec:limit}

Our evaluation brought forward some limitations of our approach, opening opportunities for future improvements. 
Some limitations are inherited from the employed computer graphics solutions, while others were introduced by the specifics of our implementation.
\rv{The papermesh after the mesh approximation should be \textit{quasi-convex and manifold}.}
This implies that the papermeshes cannot self-intersect, as they need to fit together in a physicalization. 
If \rv{intersection} happens after the mesh approximation, we have to slightly scale or translate the meshes to avoid this situation. 
Additionally, only \textit{manifold meshes with genus $0$} are currently supported. 
Whether a mesh with a genus higher than $0$ can always be unfolded into a single connected patch remains a general open topic.
\rv{As an alternative option, we can cut meshes with higher genus values to a set of meshes with genus $0$ and apply our approach.} 
Note that the shapes of the approximated meshes influences whether an inner mesh can be \textit{taken out of} its outer level.
\rv{A precise shape analysis scheme could be used to optimally cut the branches of a mesh, instead.}
It remains to be investigated if a stable solution can be reached---and even, if a possible solution can be found---with an increasing number of constraints. 
For the same reason, \textit{simple implicit functions} are used, and more complex configurations should be investigated in the future. 
\rv{Another straightforward solution would require to employ a user-defined cutting plane.}

\rv{Our initial} user study (Section~\ref{ssec:feasibility}) also showed that the combination of concave and convex patches on the meshes can lead to \textit{difficulties during the folding process}. 
\rv{The assembly of the head papercraft employed in the user study was in average 1.5 hour, 
which can is a considerable amount of time of classroom education.}
\chan{The construction and assembly time of our papercraft is reasonable based on our discussion with a professional designer~\cite{Chan:2022}. 
As also mentioned in Section~\ref{sec:intro}, a simple 3D model with $66$ polygonal faces requires a week for the creation and $10$ working days for the generation of the corresponding papercraft, while assembling it requires one hour.}
\rv{Our early investigations on advanced guidance mechanisms (Section~\ref{ssec:mountain-valley-folds}) indicate ways to simplify the construction, but we will pursue this further to identify adequate ways of reducing the user's overhead (e.g., assembly order, paper size).}


\vspace{-5pt}
\section{Conclusion and Future Work}
\label{sec:conclude}

Our paper presents a novel workflow to support the computer-aided generation of paper-based physicalizations.
The nature of nested papercrafts allows us to map anatomical or biological models to submeshes that physically represent their hierarchical topology.
The approach combines playfulness with insight, by supporting the easy design of papercrafts and data navigation through a combination of unfolding and texture projections that can be explored through colored filters or light. 
\rv{As shown in our user study, the constructed papercrafts can be used for edutainment purposes.
It would be interesting to evaluate whether this could be also used outside of the domain of anatomy or biology.}
As most prominent future improvements, a constrained-based cutting algorithm can be investigated to minimize the number of cutting planes, while extending our input to non-manifold meshes remains an open and challenging problem.
When thinking about human factors, an interesting direction would be to consider machine-assisted recommendations in the workflow in order to ease the papercraft construction process.
\rv{Finally, an evaluation that covers the entire fabrication process (and potential ways of supporting the users better, e.g., with different papers or different scorings) needs to be conducted.}

\noindent \textbf{Acknowledgments}
\vspace{-3pt}

\noindent \chan{We thank Amao Chan~\cite{Chan:2022} for sharing his knowledge on papercraft design and production, and his thoughts on future challenges.}


\bibliographystyle{eg-alpha-doi}
\bibliography{paper}



\end{document}